\pdfoutput=1
\newlength{\absize}
\documentclass[12pt]{article}
\usepackage[T1]{fontenc}
\usepackage{graphicx,dsfont,subfig,amsmath,amssymb,setspace,sectsty,float,color,cite}
\usepackage[ocgcolorlinks]{hyperref}
\hypersetup{linktocpage}
\hypersetup{linkcolor=blue,citecolor=blue,urlcolor=blue}
\renewcommand{\baselinestretch}{1.5}

\sectionfont{\Large}
\numberwithin{equation}{section}
\DeclareMathOperator{\sech}{sech}
\DeclareMathOperator{\csch}{csch}
\begin{document}
\thispagestyle{empty}
\pagestyle{empty}
\renewcommand{\thefootnote}{\fnsymbol{footnote}}
\newcommand{\starttext}{\newpage\normalsize
\pagestyle{plain}
\setlength{\baselineskip}{3.5ex}\par
\setcounter{footnote}{0}
\renewcommand{\thefootnote}{\arabic{footnote}}}
\newcommand{\preprint}[1]{\begin{flushright}
\setlength{\baselineskip}{3ex}#1\end{flushright}}
\renewcommand{\title}[1]{\begin{center}\Large\bf
#1\end{center}\par}
\renewcommand{\author}[1]{\vspace{2ex}{\normalsize\begin{center}
\setlength{\baselineskip}{3.25ex}#1\par\end{center}}}
\renewcommand{\thanks}[1]{\footnote{#1}}
\renewcommand{\abstract}[1]{\vspace{2ex}\small\begin{center}
\centerline{\bf Abstract}\par\vspace{2ex}\parbox{\absize}{#1
\setlength{\baselineskip}{3.25ex}\par}
\end{center}}
\setcounter{bottomnumber}{2}
\setcounter{topnumber}{3}
\setcounter{totalnumber}{4}
\renewcommand{\bottomfraction}{1}
\renewcommand{\topfraction}{1}
\renewcommand{\textfraction}{0}
\def\draft{
\renewcommand{\label}[1]{{\quad[\sf ##1]}}
\renewcommand{\ref}[1]{{[\sf ##1]}}
\renewenvironment{equation}{$$}{$$}
\renewenvironment{thebibliography}{\section*{References}}{}
\renewcommand{\cite}[1]{{\sf[##1]}}
\renewcommand{\bibitem}[1]{\par\noindent{\sf[##1]}}
}
\def\theequation{\thesection.\arabic{equation}}
\preprint{}
\newcommand{\be}{\begin{equation}}
\newcommand{\ee}{\end{equation}}
\newcommand{\ba}{\begin{eqnarray}}
\newcommand{\ea}{\end{eqnarray}}
\newcommand{\bas}{\begin{eqnarray*}}
\newcommand{\eas}{\end{eqnarray*}}
\newcommand{\bc}{\begin{center}}
\newcommand{\ec}{\end{center}}
\newcommand{\nn}{\nonumber}
\newcommand{\comment}[1]{}

\title{Cosmology on Compact and Stable\\ Supergravity Background}
\author{Girma Hailu\thanks{hailu@physics.harvard.edu}\\
\vspace{2ex}
\it{Jefferson Physical Laboratory\\
Harvard University\\
Cambridge, MA 02138} }

\abstract{
We propose a cosmological model of D3-brane universe on compact and stable supergravity background of wrapped D7-branes in type IIB string theory previously argued to be dual to pure $\mathcal{N}=1$ $SU(N)$ gauge theory in four dimensions. A model universe of order Planck size near the UV boundary dynamically flows toward the IR with constant total energy density and accelerating expansion followed by smooth transition to decelerating expansion and collides with the wrapped D7-branes at the IR boundary. The model addresses the horizon and flatness problems with most of the expansion produced during the decelerating expansion phase. The inflationary scenario is used to generate sources of inhomogeneities in the cosmic microwave background radiation and seeds for large scale structure formation from quantum fluctuations which exit the Hubble radius early during the accelerating expansion phase and the model addresses the inhomogeneity problem with red tilt in the power spectrum. We propose that the kinetic energy of the model universe is converted to matter and radiation by the collision followed by formation of baryons that stabilizes the model universe against gravitational force from the background at a finite distance from the IR boundary with the wrapped D7-branes serving as sources of color. Friedmann evolution then takes over with a positive cosmological constant term coming from the remaining potential energy density which is interpreted as dark energy. The magnitude of dark energy density is smaller than the total energy density during the flow by a ratio of the scale factor when the model universe appears in the UV to the scale factor at the moment of collision and stays constant while the matter-radiation density falls during Friedmann expansion.
}

\starttext
\newpage
\tableofcontents

\section{\label{sec:intro}Introduction}

One important experimental observation in early $20^{th}$ century cosmology was that the universe was not static but expanding \cite{Hubble:1929ig}, and according to
the hot big bang hypothesis the universe originated with sub-Planckian size and expanded to its current size. Two of the limitations of the old hot big bang model are the horizon problem (on why causally disconnected regions look nearly identical) and flatness problem (on why the universe is spatially nearly flat, which requires a highly fine-tuned initial rate of expansion for a given initial matter distribution).

The inflationary paradigm \cite{Guth:1980zm, Linde:1981mu, Albrecht:1982wi} addresses these limitations by positing that the universe underwent a period of accelerated expansion whereby an initially Planck size universe gets exponentially stretched by a factor of about $10^{28}$ or more which was then followed by Friedmann evolution of the old big bang model.
In standard inflationary models, inflation corresponds to the motion of a scalar inflaton field in a given potential, and a suitable inflaton potential is introduced by hand and its consequences are investigated which, in addition to resolving the limitations of the old big bang model, serve as a probe to fundamental physics at high energies through current cosmological observations of the early universe. At the end of inflation, the scalar field is believed to decay to particles which fill the early universe in a process called reheating. As the universe expanded and cooled, baryons, atoms, nuclei, and  then later large scale structures whose origins are quantum fluctuations during inflation formed.  The same quantum fluctuations are sources of thermal fluctuations in the cosmic microwave background radiation (CMBR). The universe is nearly homogeneous and isotropic with thermal fluctuations in the CMBR being of order $10^{-5}$. Details in the power spectrum together with required magnitude of inflation needed to address the horizon and flatness problems put constraints on the inflaton potential.

A second more recent important cosmological observation was that the expansion of the universe is currently accelerating \cite{Perlmutter:1998np}, which is attributed to the presence of dark energy accounting for about $70\%$ of the total energy content of the universe.

Flux compactifications in string theory provide a scheme for constructing models of the universe in which the cosmic evolution could be dynamically determined for a given background. A brane inflation model in which the universe is a D3-brane and the inflaton field is the position of the D3-brane moving due to interaction with an anti-D3-brane was proposed in \cite{Dvali:1998pa} and further investigation on stabilized supergravity background was initiated in \cite{Kachru:2003aw}.

A duality between $N$ D7-branes filling four-dimensional spacetime and wrapping a 4-cycle on $\mathds{R}^{1,3}\times \frac{\mathds{C}^1}{Z_2}\times \frac{\mathds{T}^2}{Z_2}\times \frac{\mathds{T}^2}{Z_2}$ and  pure $\mathcal{N}=1$ $SU(N)$ gauge theory in four dimensions was proposed and argued in \cite{Hailu:2011pn}. The wrapped D7-branes turn on all  $F_1$, $F_3$, $H_3$, and $F_5$ fluxes of type IIB theory and induce torsion.
The supergravity solutions were explicitly constructed with exact analytic expressions for all components of the metric and fluxes.
The background is compact and stable as a result of supersymmetry and a balance between fluxes and torsion. The geometry is complex, non-K\"ahler, and conformally Calabi-Yau. The wrapped D7-branes are located at the IR boundary where there are five internal angular directions. The induced fluxes warp space and result in four-dimensional spacetime at the UV boundary located at a finite distance from the IR boundary. The IR and UV boundaries are respectively located at $r=r_s$ and $r=r_s\, e^{\frac{2\pi}{3g_sN}}$ on the radial direction $\mathds{R}^1$ in $\frac{\mathds{C}^1}{Z_2}\sim \mathds{S}^1\times \mathds{R}^1$, where $r_s$ is the radius of the angular directions at the IR boundary and $g_sN$ is the 't Hooft coupling at the UV boundary.
It was shown in \cite{Hailu:2011pn} that the supergravity background reproduces the renormalization group flow and pattern of chiral symmetry breaking of the gauge theory and in \cite{Hailu:2011kp} that it produces linear confinement of quarks within consistent relativistic quantum theory of ten-dimensional superstrings for the first time. In \cite{Hailu:2012jc} mass spectrum of $0^{++}$ glueballs in $\mathcal{N}=1$ $SU(N)$ gauge theory in the large $N$ limit was produced whose ratios happen to agree with available data from large $N$ nonsupersymmetric lattice QCD to within $5\%$, which is consistent with linear confinement in both $\mathcal{N}=1$ supersymmetric and nonsupersymmetric $SU(N)$ gauge theories.

In this paper, we argue that the fact that the evolution of the universe cannot be accounted by its visible matter and energy content might partly be due to forces external to it and propose a cosmological model of D3-brane universe on the supergravity background in \cite{Hailu:2011pn} with the following assumptions:
\begin{itemize}
\item A D3-brane universe of order Planck size filling four-dimensional spacetime $\mathds{R}^{1,3}$ spontaneously appears in the UV region.
\item When the D3-brane collides with the wrapped D7-branes at the IR boundary, its kinetic energy is converted to matter and radiation confined to the D3-brane.
\end{itemize}
The first assumption on spontaneous appearance of the model universe is similar to the case in the standard or other cosmological models where the origin or existence of a universe is always assumed. The second assumption is analogous to the process of reheating whose details are not well understood and is currently an assumption in inflationary cosmology models.

The dynamics of the D3-brane until collision is governed by the Dirac-Born-Infeld (DBI) action that is  determined by the background.
The evolution of the D3-brane is analyzed starting with one located near the UV boundary. For examples of cosmological models and analyses  which make use of the DBI action, see \cite{Kachru:2003sx,Silverstein:2003hf,Alishahiha:2004eh,Chen:2005fe,Shandera:2006ax,
Kecskemeti:2006cg,Hailu:2006uj,Bean:2007eh,Bean:2008na,Firouzjahi:2010ga,Miranda:2012rm} for instance. The background we have here is compact and it does not involve $AdS_5$ space since it is related to a confining gauge theory that is scale dependent. Compactness of the background geometry, availability of wrapped D7-branes at the IR boundary, and confinement of quarks that it produces provide a convenient setting for the beginning, reheating, and final stabilization of the model universe which then follows Friedmann evolution and it does not need to be glued to a larger background. Moreover, for $N>>1$ string loop corrections are negligible and the background is not measurably
perturbed by the introduction of a single D3-brane. For small enough $g_sN$, the curvature is small and $\alpha'$ corrections are negligible.

It is found that the background geometry is such that the D3-brane flows toward the IR with accelerating expansion of its spatial volume followed by smooth transition to decelerating expansion. The cosmic evolution is dynamically dictated by the warped background geometry of the wrapped D7-branes with no additional branes or potentials introduced by hand.
The total energy density of the D3-brane stays constant with
the initial potential energy density getting converted to kinetic energy density as it flows toward the IR. Because the background preserves $\mathcal{N}=1$ supersymmetry in four dimensions, both the wrapped D7-branes and the probe D3-brane stay flat on $R^{1,3}$.
The magnitude of the scale factor of the three-dimensional volume produced is large enough to address the horizon and flatness problems provided that the D3-brane starts from close enough to the UV boundary.

The D3-brane reaches the IR boundary with velocity approaching its speed limit and collides with the wrapped D7-branes.
We propose that the kinetic energy of the model universe is converted to matter and radiation by the collision followed by formation of baryons with strong interaction modeled by strings between quarks and the wrapped D7-branes serving as sources of color.
The separation between quarks in baryons does not grow as the universe expands which gives a corresponding finite distance between the strings which end on quarks on the D3-brane and the D7-branes at the IR boundary, and strong interaction stabilizes the model universe at a finite distance from the wrapped D7-branes against gravitational force from the background.
Friedmann evolution then takes over with a positive cosmological term coming from the remaining potential energy density of the model universe in the background which serves as a source of dark energy.
The magnitude of dark energy density is smaller than the total energy density during the flow by a factor of the ratio of the scale factor when the model universe appears in the UV to the scale factor at the moment of collision at the IR boundary and stays constant while the matter-radiation density falls during Friedmann expansion, and it produces accelerating expansion during dark energy dominated phase.

The inflationary scenario is used to generate sources of thermal fluctuations in the CMBR and seeds for large scale structure formation from quantum fluctuations which exit the Hubble radius early during the accelerating expansion phase, and the model addresses the inhomogeneity problem. The power spectrum has red tilt consistent with observations.
We show that slow-roll inflation can be used in this very early phase of cosmic evolution. Using results form slow-roll inflation and constraints from experimental cosmological observations on density perturbations, spectral tilt, and dark energy together with measured value of Newton's gravitational constant in four dimensions, we write down constraints on the parameters of the model.

We then consider one example of numerical values of the parameters which satisfy all the constraints except that the dark energy density is not small enough (but reduced from Planck scale in the right direction) and discuss their physical interpretations. Further investigation of the model, its parameters space, and available cosmological data is suggested considering the importance of addressing these problems simultaneously in the same setting which also provides a dual gravity theory for investigating the nonperturbative dynamics of $\mathcal{N}=1$ supersymmetric $SU(N)$ gauge theory. We also discuss how the dynamics might be used to construct a cyclic universe.

\section{Dynamics of D3-brane\label{sec:eqnofm}}

In this section, we write the DBI action for a probe D3-brane on the supergravity background of \cite{Hailu:2011pn} and analyze its dynamics. First the ten-dimensional  metric is
\be
ds_{10}^2={r_s^2}\,\Bigl({\cosh u}\,g_{\mu \nu}dx^\mu dx^\nu
+ \sech u\,\left({d \rho^2}+d\psi^2+d\varphi_1^2+d\varphi_2^2
+d\varphi_3^2+d\varphi_4^2\right)\Bigr),\label{metric10-1a} 
\ee
where
\be
\sech u= \sqrt{1-(\frac{g_sN}{2\pi}\rho)^2},\,\qquad \rho = 3 \ln{r},\label{u-rho-1}
\ee
$x^\mu=(\tau, \vec{x})$ denotes the coordinates on four-dimensional spacetime with metric $g_{\mu\nu}$ having $(-,+,+,+)$ signature and $\tau$ being time, $(\psi,\varphi_1,$ $\varphi_2,\varphi_3, \varphi_4)$ are angular coordinates on the compact transverse internal space, and $r$ is the radial coordinate. All components of $x^\mu$ and $r$ are measured in units of $r_s$, the radius of the angular directions at the IR boundary.\footnote{This notation avoids factors of $r_s$ floating around in our calculations and physical quantities are represented by dimensionless variables in units of $r_s$ to appropriate power with $c=\hbar=1$. We will put $r_s$ back and restore the dimensions to all variables, again in $c=\hbar=1$ units, starting from the last paragraph of section \ref{sec:hf} where we begin testing the model quantitatively with experimental data.}
The coordinates in ten dimensions and the corresponding metric will be denoted by $X^M$ and $G_{MN}$. The dilaton $\Phi$ is constant and the string coupling is given by $e^\Phi=g_s$.
The IR boundary is at $r=1$ or $\rho=0$ and the UV boundary is at $r=e^{\frac{2\pi}{3g_sN}}$ or $\rho=\frac{2\pi}{g_sN}$, where $g_s N$ is the 't Hooft coupling at the UV boundary. The metric together with the fluxes on the background (all $F_1$, $F_3$, $H_3$, and $F_5$ fluxes of type IIB theory are turned on) solves all the supergravity equations of motion. See \cite{Hailu:2011pn} for details on the supergravity background.

\subsection{Dirac-Born-Infeld action\label{sec:action}}

We want to study the dynamics of a probe D3-brane that fills four-dimensional spacetime and whose radial location on the background is represented by $\rho=\rho(x^\mu)$. The metric given by (\ref{metric10-1a}) then becomes
\ba
ds_{10}^2={r_s^2}\,\Bigl({\cosh u}(g_{\mu\nu}+\partial_\mu\rho\, \partial_\nu\rho\,\sech^2 u)\,{dx^\mu dx^\nu}\nn\\
+ \sech u\,\left(d\psi^2+d\varphi_1^2+d\varphi_2^2
+d\varphi_3^2+d\varphi_4^2\right)\Bigr).\label{metric10-2a}
\ea

The DBI action for the D3-brane is
\be
S_{\mathrm{DBI}}=-\frac{T_3}{r_s^4}\int d^4x\, e^{-\Phi} \sqrt{-\det G_{\mu\nu}}\,,\label{DBI-D3-1}
\ee
where 
\be
G_{\mu\nu}={r_s^2}\,{\cosh u}(g_{\mu\nu}+\partial_\mu\rho\, \partial_\nu\rho\,\sech^2 u)\label{Gmunu-induces-1}
\ee
is the induced metric on the worldvolume and $T_3$ is the tension of the D3-brane, measured in units of $r_s^{-4}$. Because there is no $C_4$ potential in the background, there is no corresponding Chern-Simons term in (\ref{DBI-D3-1}). With (\ref{Gmunu-induces-1}), the action becomes
\be
S_{\mathrm{DBI}}=-\frac{T_3}{g_s}\int  d^4x\,\cosh^2 u \,\sqrt{-\det (g_{\mu\nu}+\partial_\mu\rho\, \partial_\nu\rho\sech^2 u)}.\label{DBI-D3-1a0}
\ee
For the radial time evolution of the D3-brane excluding perturbations, which come in when we study inhomogeneities, $\rho=\rho(\tau)$ and the action reduces to
\be
S_{\mathrm{DBI}}=-\frac{T_3}{g_s}\int  d^4x\,\cosh^2 u \,\sqrt{1- \dot{\rho}^2\sech^2 u}.\label{DBI-D3-1a}
\ee
Extremizing (\ref{DBI-D3-1a}), the equation of motion of the D3-brane is
\be
\ddot{\rho}=-(\frac{g_sN}{2\pi})^2 \rho\,\cosh^4 u \left(2- 3\dot{\rho}^2\,\sech^2 u \right).\label{rho-eqn-1}
\ee

Rewriting the metric given by (\ref{metric10-2a}), now with $\rho=\rho(\tau)$, as
\ba
ds_{10}^2={r_s^2}\,{\cosh u}\,(1-\dot{\rho}^2\sech^2 u)\,\left(-{d\tau}^2+\,a^2\,d\vec{x}_{3}^2\right)\nn\\
+ {r_s^2}\,\sech u\,\left(d\psi^2+d\varphi_1^2+d\varphi_2^2
+d\varphi_3^2+d\varphi_4^2\right),\label{metric10-2b}
\ea
we see that the scale factor for the three-dimensional space is given by
\be
a=\frac{1}{\sqrt{1-\dot{\rho}^2\sech^2 u}}.\label{a-1}
\ee

Consider a D3-brane that is located near the UV boundary at $\rho(\tau_i)=\frac{2\pi}{g_sN}-\delta$ at time $\tau_i$, where $\delta<<\frac{2\pi}{g_sN}$.
We see from (\ref{rho-eqn-1}) and (\ref{u-rho-1}) that the D3-brane flows toward the IR boundary. Its initial acceleration is large provided that the initial location is close enough to the UV boundary, since $\cosh u>>1$, and the evolution of the D3-brane is not sensitive to its initial speed which we set to zero, $\dot{\rho}(\tau_i)=0$.
The scale factor given by (\ref{a-1}) increases as the D3-brane flows toward the IR with the value of $\dot{\rho}$ getting close to its speed limit of $-1$ and $\sech u=1$ at the IR boundary. The magnitude of the scale factor when the D3-brane reaches the IR boundary can be made as large as desired by taking $g_sN\,\delta$ to be appropriately small. In the limit $g_sN\,\delta\to0$, the scale factor becomes infinitely large. The corresponding volume of the three-dimensional space grows by a factor of order $(\frac{\pi}{g_sN\, \delta})^3$.
The D3-brane then collides with the stack of $N$ wrapped D7-branes when it reaches the IR boundary at $\rho=0$.

\begin{figure}[t]
\centering
\subfloat{\label{fig:rho3}\includegraphics[width=0.41\textwidth]
{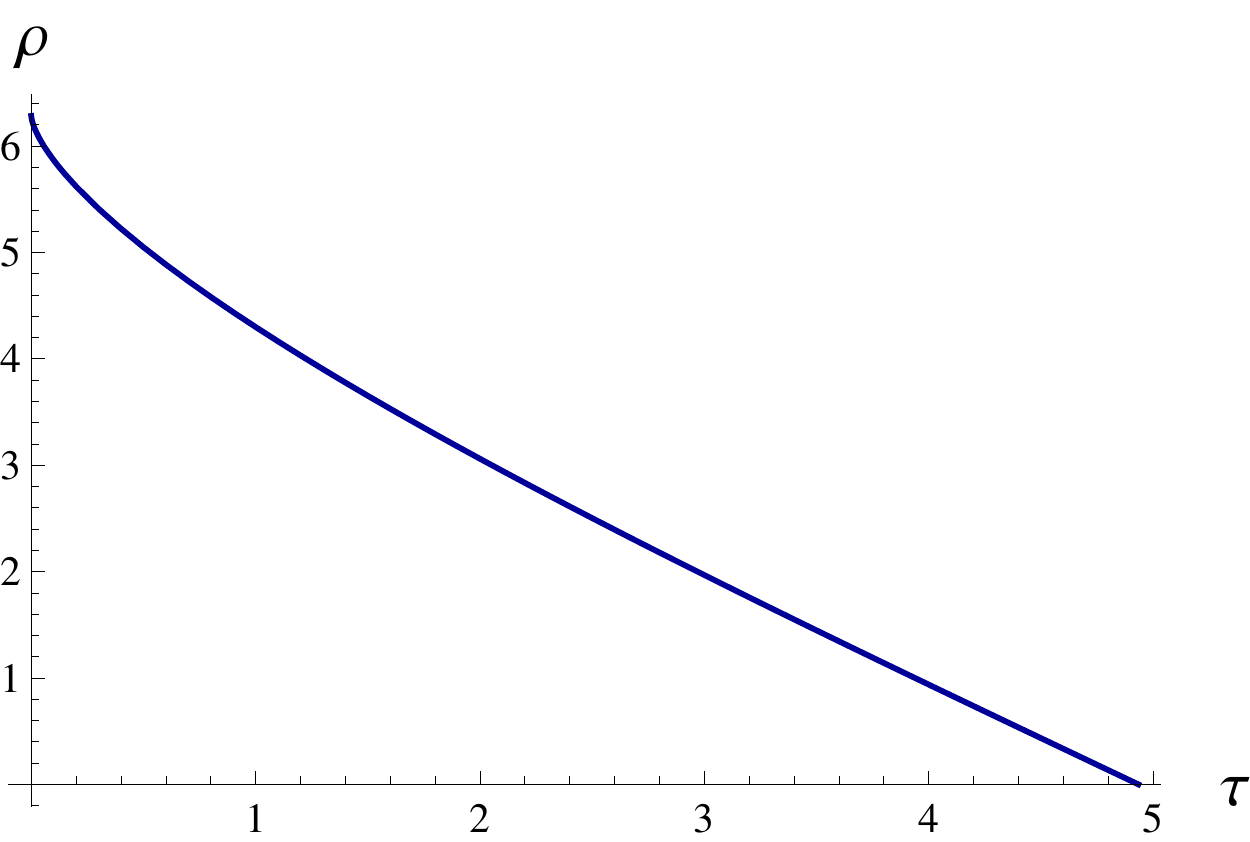}}
\hspace{15.0mm}
\subfloat{\label{fig:a}\includegraphics[width=0.41\textwidth]
{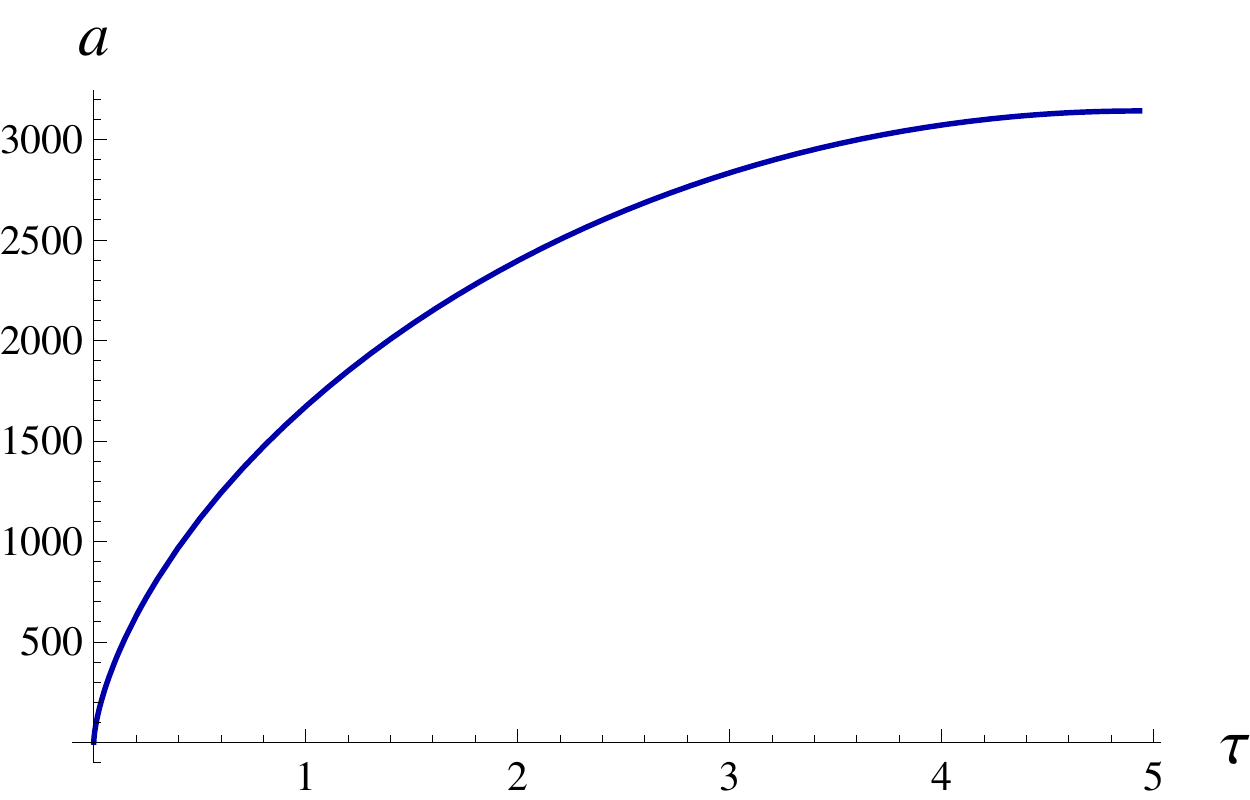}}
\hspace{5mm}
\subfloat{\label{fig:rho3n}\includegraphics[width=0.44\textwidth] 
{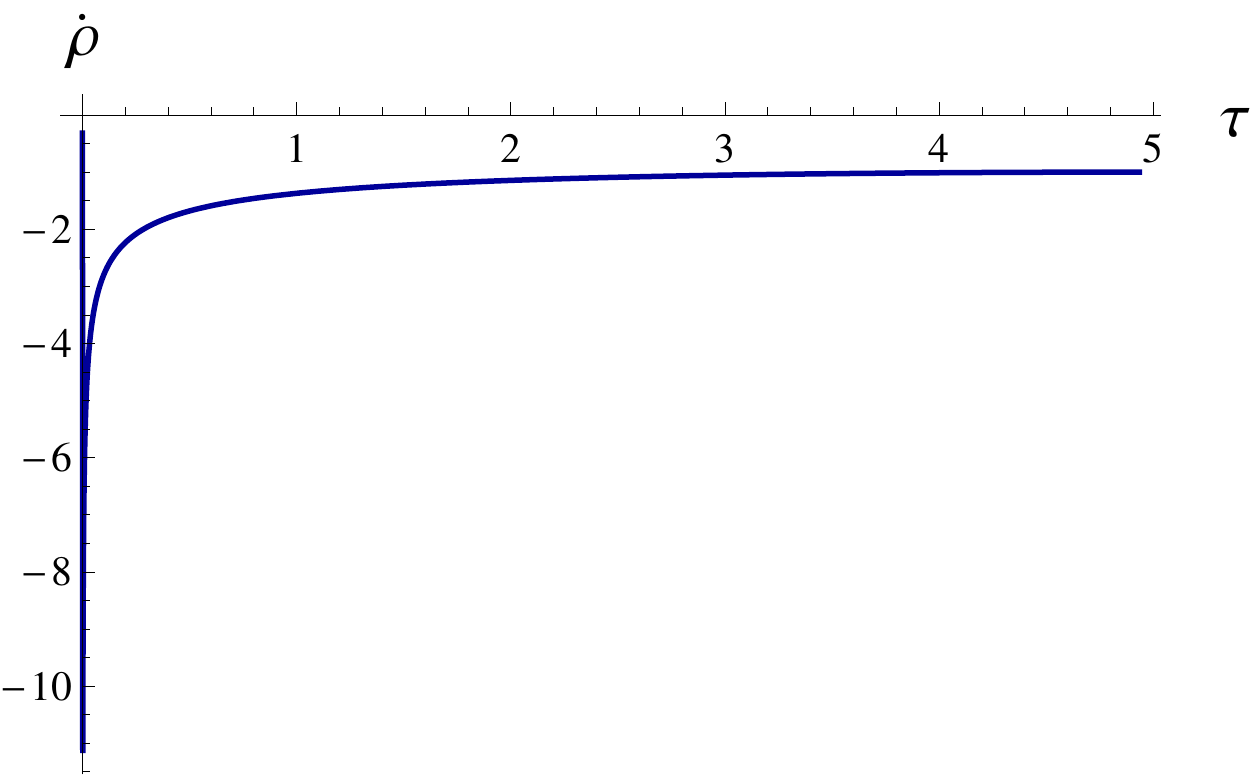}}
\hspace{11.5mm}
\subfloat{\label{fig:ad3n}\includegraphics[width=0.44\textwidth]
{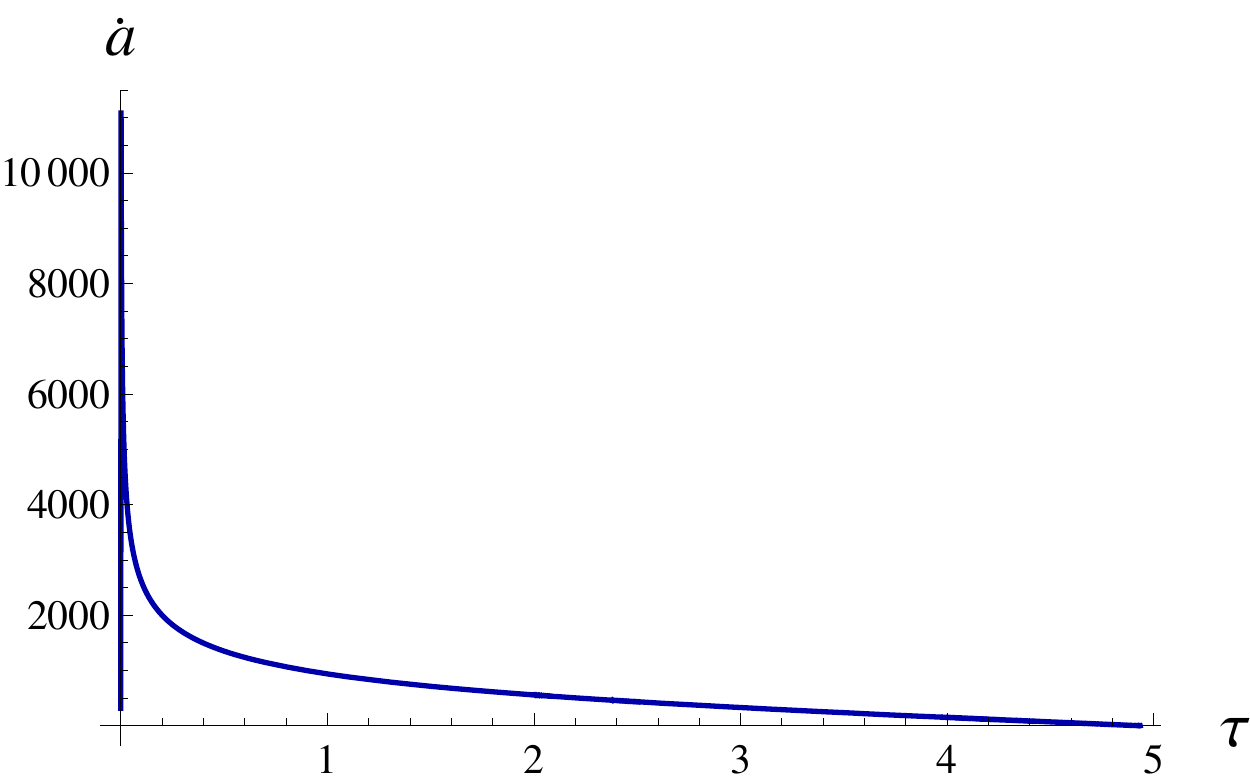}}
\hspace{5mm}
\subfloat{\label{fig:rho3ds}\includegraphics[width=0.44\textwidth]
{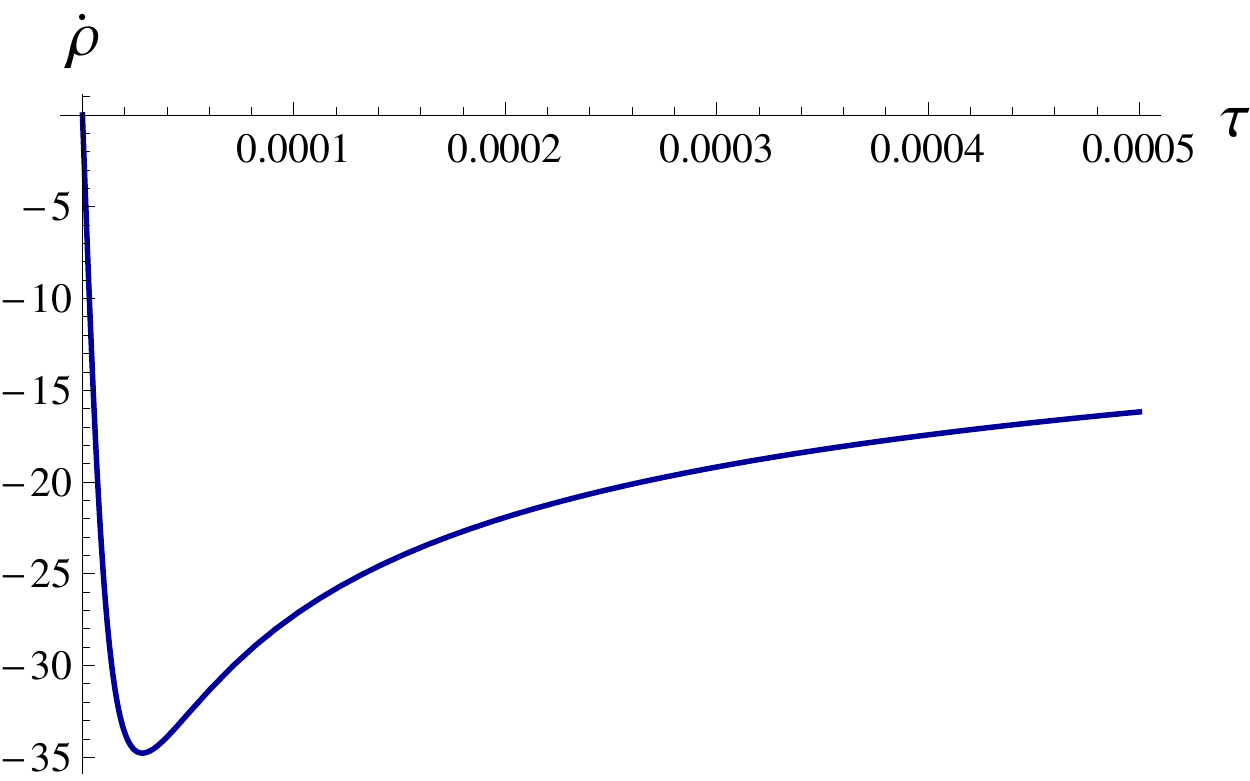}}
\hspace{11.75mm}
\subfloat{\label{fig:ads}\includegraphics[width=0.44\textwidth]
{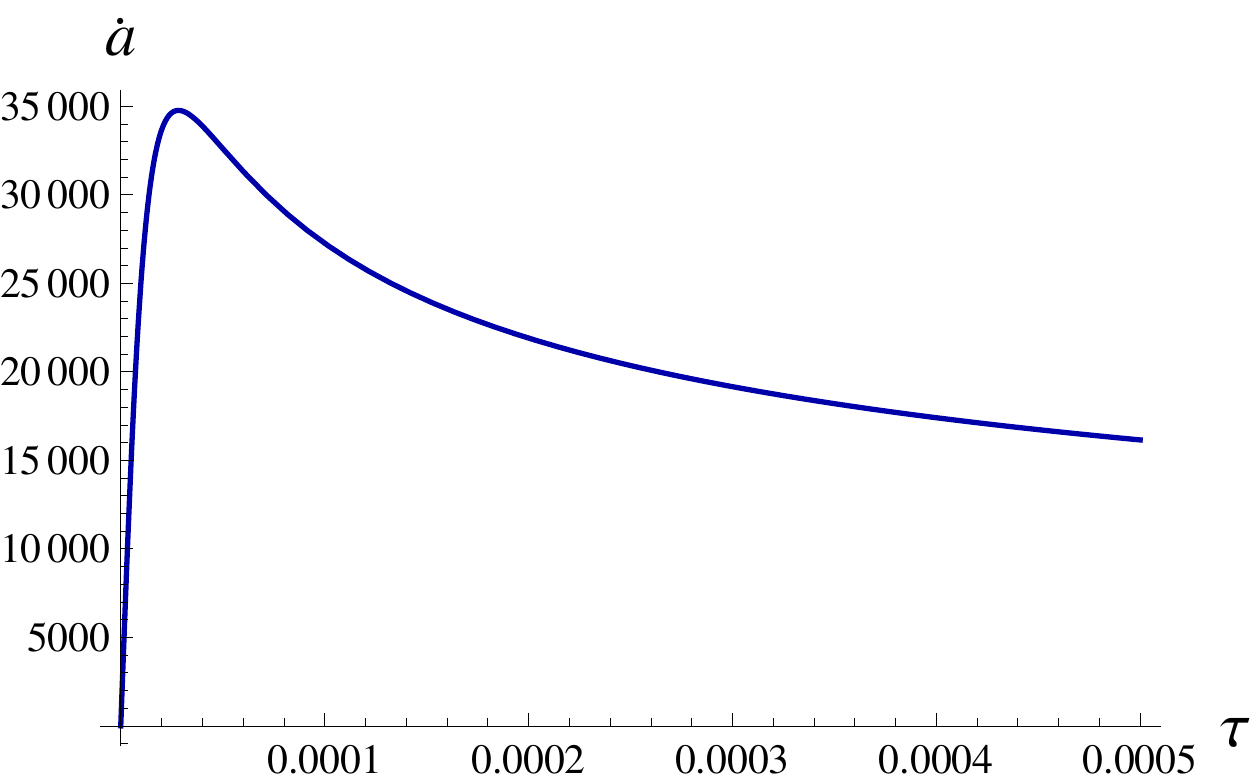}}
\hspace{2mm}
\subfloat{\label{fig:rho3dds}\includegraphics[width=0.48\textwidth]
{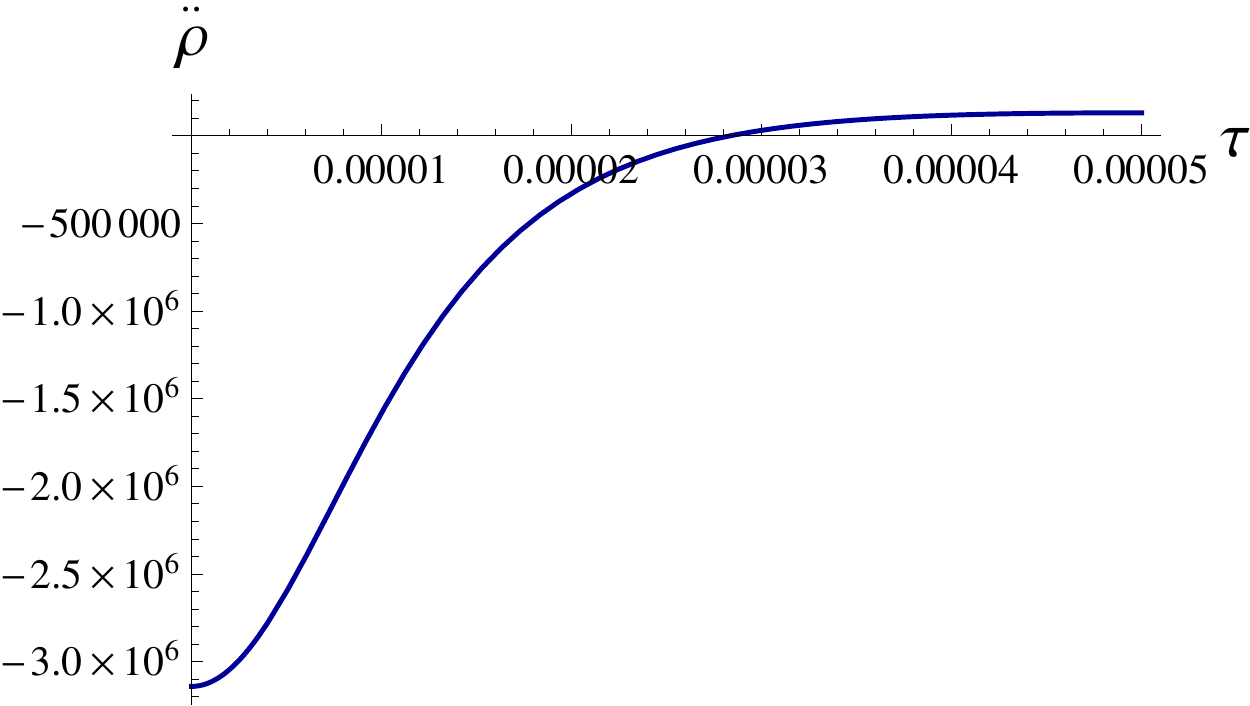}}
\hspace{6mm}
\subfloat{\label{fig:adds}\includegraphics[width=0.47\textwidth]
{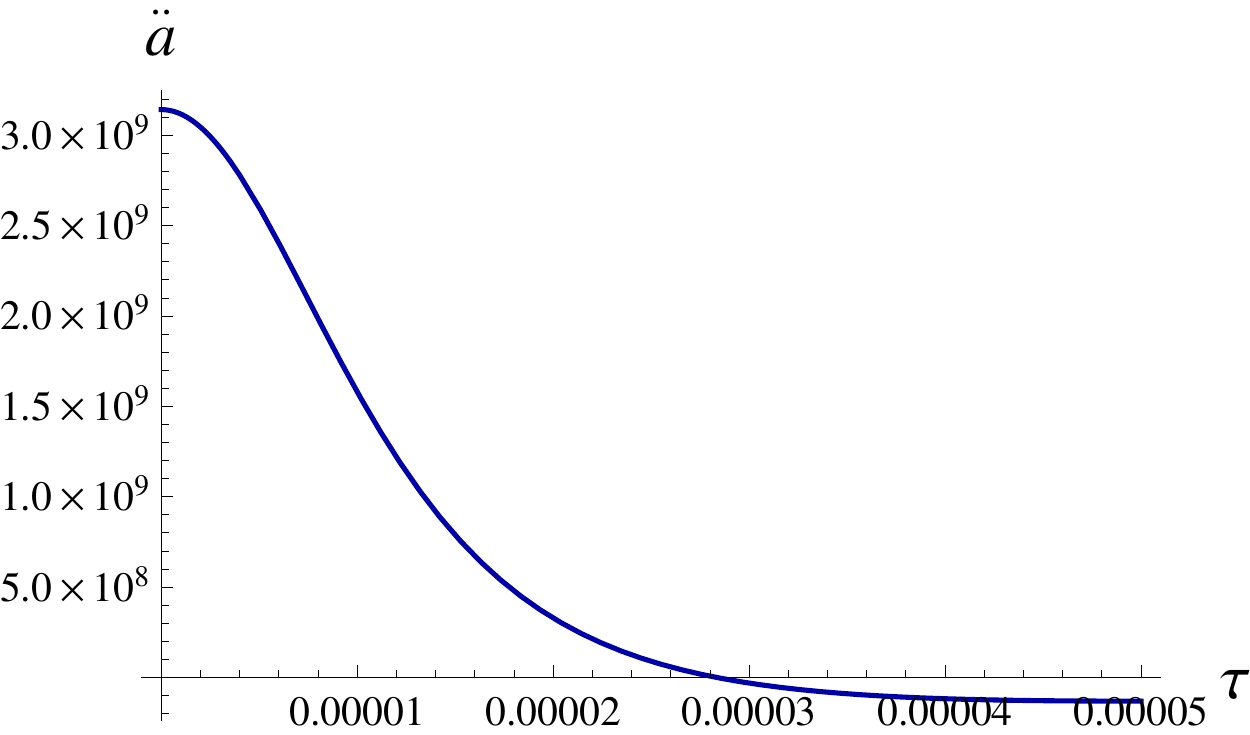}}
\caption{Time evolution of  $\rho$ and $a$ for the flow from near the UV boundary to the IR boundary and the corresponding $\dot{\rho}$,  $\dot{a}$, $\ddot{\rho}$, and $\ddot{a}$ for $g_sN=1$ and $\delta=10^{-3}$. Some of the plots are made for shorter ranges of time to show the transition from accelerating to decelerating expansion closely.}\label{figs:rho-a-3}
\end{figure}

Numerical solutions to (\ref{rho-eqn-1}) are shown in figure \ref{figs:rho-a-3}
for the time evolution of $\rho$ and $a$ and their derivatives with $\rho(\tau_i)=\frac{2\pi}{g_sN}-\delta$, $\dot{\rho}(\tau_i)=0$, and the initial time $\tau_i$ set to zero for $g_sN=1$ and $\delta=10^{-3}$, with the value of the parameters chosen only for illustration of the features. We will write analytic expressions in section \ref{sec:Ham}. Notice that the accelerating expansion takes place in a small fraction of the time interval during which the scale factor does not grow much. The speed of expansion then decreases while still staying large ($>>1$) for nearly all of the flow for the parameters in the plot, and it is during this decelerating expansion phase that most of the growth of the scale factor occurs.
The whole evolution is dynamically generated by the DBI action which is determined by the background; we have not introduced any additional terms to the equation of motion by hand. 

\emph{Thus the supergravity background produces dynamical accelerating expansion of the D3-brane followed by smooth transition to decelerating expansion with no additional ingredients introduced by hand, and the scale factor becomes larger as the initial location of the D3-brane gets closer to the UV boundary for a given $g_sN$.}

Finally, the equation of motion of the D3-brane given by (\ref{rho-eqn-1}) can also be rewritten in different but equivalent forms such as
\be
\ddot{u}=-(2(\frac{g_sN}{2\pi})^2 \cosh^6 u -5\dot{u}^2 ) \tanh u,\label{u-eqn-1c}
\ee
which involves only $u$, and
\be
\frac{d}{d\tau}\left(\dot{\rho}\,a\right)=-\frac{g_sN}{2\pi}\cosh^3u\sinh u\,\left(a+\frac{1}{a}\right).\label{rho-eqn-1c}
\ee

\subsection{Energy density\label{sec:Ham}}

The energy (or Hamiltonian) density for the D3-brane corresponding to the DBI action given by (\ref{DBI-D3-1a}) is
\be \mathcal{H}=\frac{T_3}{g_s}\,\frac{\cosh^2 u}{\sqrt{1- \dot{\rho}^2\sech^2 u}}\,=\,\sqrt{p_\rho^2\,\cosh^2 u+(\frac{T_3}{g_s})^2\,\cosh^4u}\,,\label{H-D3-1a}
\ee
where $p_\rho$ is the conjugate momentum to $\rho$,
\be
p_\rho=\frac{T_3}{g_s}\,\frac{\dot{\rho}}{\sqrt{1- \dot{\rho}^2\sech^2 u}}\,.
\ee
The potential and kinetic energy densities of the D3-brane are respectively given by
\be
\mathcal{V}=\frac{T_3}{g_s}\,\cosh^2u\label{V-1}
\ee
and
\be
\mathcal{T}=\frac{T_3}{g_s}\,\frac{\cosh^2 u}{\sqrt{1- \dot{\rho}^2\sech^2 u}}-\frac{T_3}{g_s}\,\cosh^2u=\sqrt{p_\rho^2\,\cosh^2 u+(\frac{T_3}{g_s})^2\,\cosh^4u}-\frac{T_3}{g_s}\,\cosh^2u.\label{T-1}
\ee

Using the energy density given by (\ref{H-D3-1a}) and the equation of motion given by (\ref{rho-eqn-1}), we obtain
\be
\frac{d\mathcal{H}}{d\tau}=0.\label{H-const-1}
\ee
\emph{Thus the total energy density of the probe D3-brane stays constant during the flow (and expansion) between the UV and IR boundaries.}

\begin{figure}[t]
\centering
\subfloat{\label{fig:HVT1}\includegraphics[width=0.55\textwidth]
{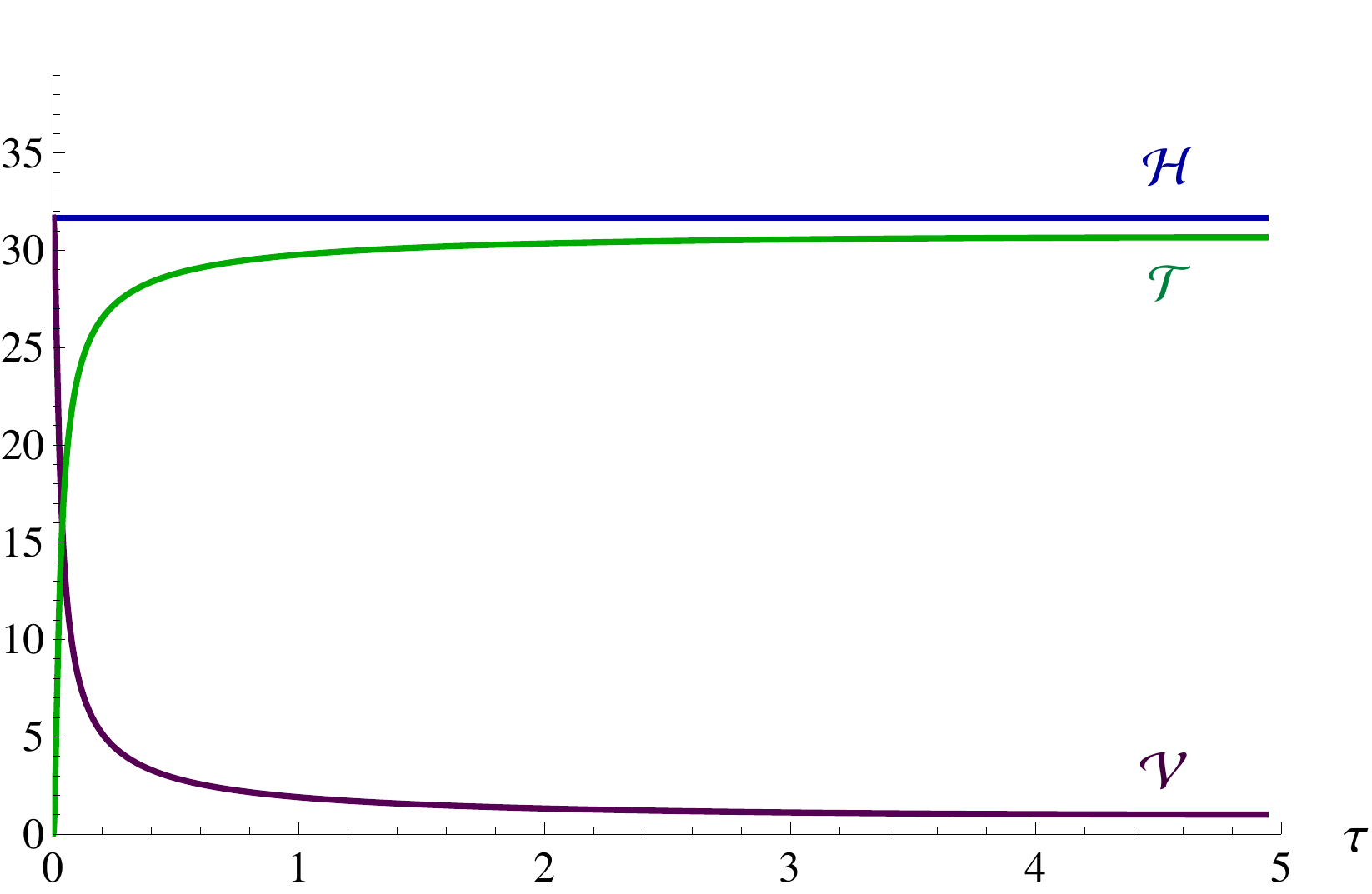}}
\\
\subfloat{\label{fig:HVT1ss}\includegraphics[width=0.55\textwidth]
{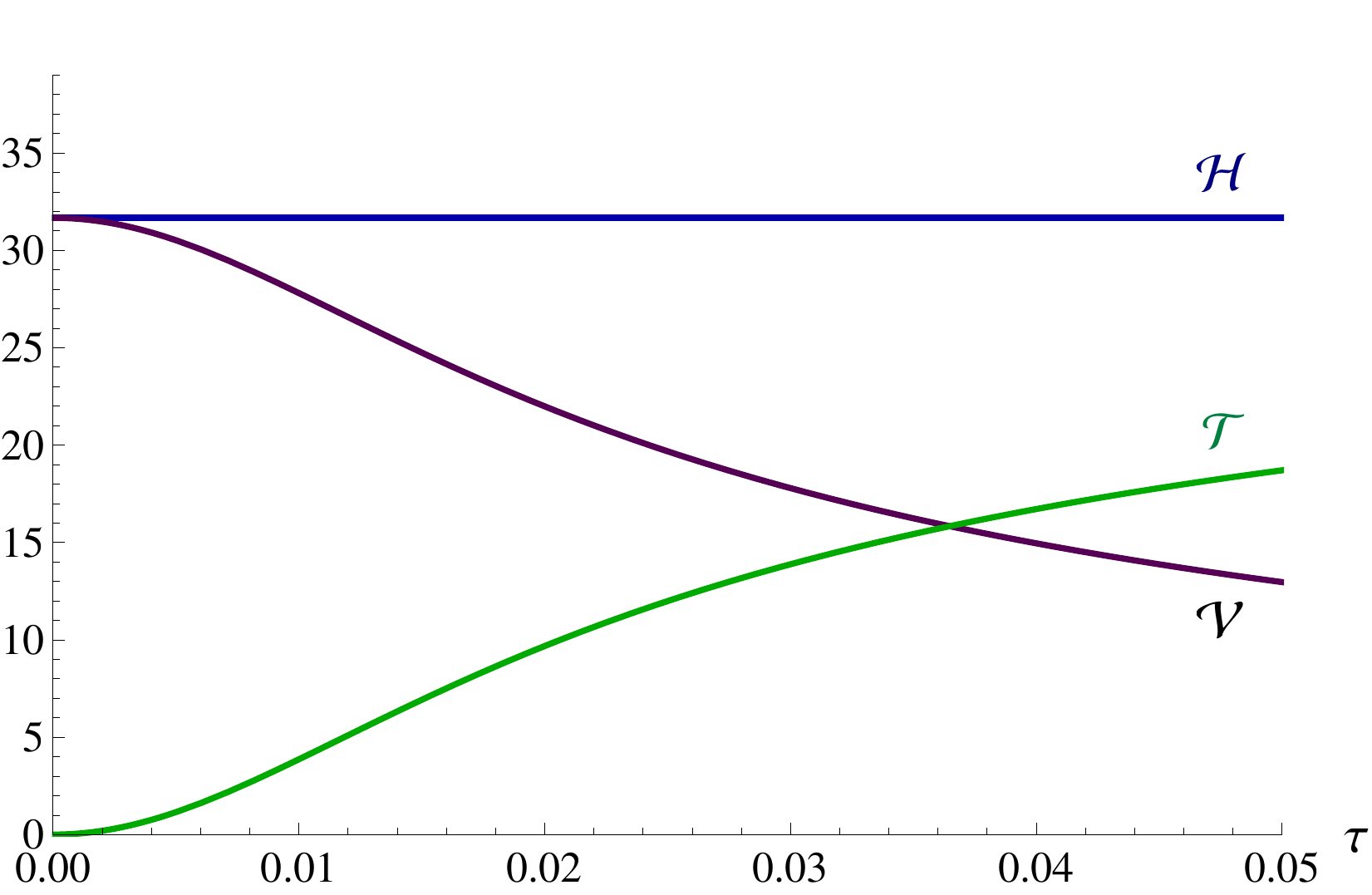}}
\caption{Total, kinetic, and potential energy densities of D3-brane measured in units of $\frac{T_3}{g_s}$ for $g_sN=1$ and $\delta=0.1$ with the first plot showing the time evolution until collision and the second one showing a shorter time range for the early phase.}\label{figs:HVT}
\end{figure}

Consider a D3-brane that starts with $\dot{\rho}(\tau_i)=0$ at ${\rho}(\tau_i)=\frac{2\pi}{g_s N}-\delta$ with $\delta<< \frac{2\pi}{g_s N}$. The energy density is $\mathcal{H}(\tau_i)=\frac{T_3}{g_s}\,{\cosh^2 u(\tau_i)}$ initially at time $\tau_i$ and $\mathcal{H}(\tau_c)=\frac{T_3}{g_s}\,\frac{1}{\sqrt{1- \dot{\rho}(\tau_c)^2}}$ when the D3-brane reaches the IR boundary at time $\tau_c$. The energy density of the D3-brane stays constant with most of the energy density which initially at time $\tau_i$ is all potential converted to kinetic energy density  at time $\tau_c$, where $\mathcal{T}\approx p_\rho$. Because we take $\delta<< \frac{2\pi}{g_s N}$, we will from now on mostly do calculations to leading order in $\frac{g_sN\,\delta}{2\pi}$ and
\be
\mathcal{H}(\tau_i)=\frac{T_3}{g_s}\,\frac{\pi}{g_s N \delta}=\mathcal{H}(\tau),\label{H-const-2}
\ee
where $\mathcal{H}(\tau)$ is the energy density at any time between $\tau_i$ and $\tau_c$. Figure \ref{figs:HVT} shows the time evolution of the energy densities for $g_sN=1$ and $\delta=0.1$, with the parameters chosen only for illustration features including the remaining potential energy density when  the D3-brane reaches the IR boundary.

Analytic expressions for the scale factor $a$ and the rate of change of ${\rho}$ at any location $\rho$  to leading order in $\frac{g_sN\,\delta}{2\pi}$ are obtained using constancy of energy density, $\mathcal{H}(\tau)=\mathcal{H}(\tau_i)$,
\be
a=\frac{\pi}{g_sN\delta}\sech^2 u\,, \qquad \dot{\rho}=-\cosh u\, \sqrt{1-(\frac{g_s N\delta}{\pi}\cosh^2u)^2}\,,
\ee
\be
\dot{a}=-\frac{1}{\delta}\,\tanh u\,\dot{\rho}=\frac{1}{\delta}\,\sinh u \,\sqrt{1-(\frac{g_s N\delta}{\pi}\cosh^2u)^2}.\label{adot-analy-1}
\ee
Notice that initially $a(\tau_i)=1$ and $\dot{\rho}(\tau_i)=0$ as we had required and
\be
 \ddot{\rho}(\tau_i)= -\frac{\pi}{g_sN\, \delta^2}\,,\qquad \ddot{a}(\tau_i)=  \frac{\pi}{g_sN\, \delta^3}.\label{ddrho-dda-1}
\ee
When the D3-brane reaches the IR boundary,
\be
a(\tau_c)= \frac{\pi}{g_sN\, \delta}\,,\qquad \dot{\rho}(\tau_c)=- \sqrt{1-(\frac{g_s N\delta}{\pi})^2}\,,\label{a-rho-ts-1}
\ee
and the scale factor can be made large by taking $g_sN\,\delta$ to  be appropriately small. The speed of the D3-brane when it collides with the wrapped D7-branes is such that $\dot{\rho}(\tau_c)\to -1$ as $g_sN\,\delta\to 0$. Because $\delta<< \frac{2\pi}{g_sN}$,  $\dot{\rho}(\tau_c)\approx -1$. In fact, the D3-brane approaches the speed limit and $\dot{\rho}(\tau)\approx -1$ for most of the flow after the transition from accelerating to decelerating expansion. Notice that the scale factor at time of collision is related to the total energy density,
\be
a(\tau_c)=\frac{g_s}{T_3} \mathcal{H}.
\ee
Thus both the scale factor and the energy density increase with decreasing $\delta$ as the initial position of the D3-brane gets closer to the UV boundary and become infinite when $\delta=0$ for a given $g_sN$. We study the evolution of the D3-brane starting from $\delta\ne 0$, the background geometry is singular at the UV boundary where the curvature diverges because of warping of the transverse internal space to zero-size.

\section{Cosmological model\label{sec:modelU}}

In this section, we use the D3-brane dynamics studied in section \ref{sec:eqnofm} to construct a cosmological model and discuss its features.

\subsection{Expansion of model universe\label{sec:infl}}

We assume that a model D3-brane universe of order Planck size spontaneously appears near the UV boundary at $\rho(\tau_i)=\frac{2\pi}{g_sN}-\delta$ at time $\tau_i$ with $\dot{\rho}(\tau_i)=0$, where $\delta<<\frac{2\pi}{g_sN}$. We have already studied the expansion in detail in section \ref{sec:eqnofm}. Our convention for the scale factor is such that $a(\tau_i)=1$. The model universe initially has potential energy and dynamically flows to the IR boundary while undergoing accelerating expansion followed by smooth transition to decelerating expansion and collides with the wrapped D7-branes at the IR boundary at time $\tau_c$ with speed $\dot{\rho}(\tau_c)=- \sqrt{1-(\frac{g_s N\delta}{\pi})^2}$ and a scale factor of $a(\tau_c)= \frac{\pi}{g_sN\, \delta}$. The flow and expansion takes place with the total energy density staying constant, $\mathcal{H}=\frac{T_3}{g_s}\,\frac{\pi}{g_s N \delta}$.

The kinetic energy of the D3-brane is converted to matter and radiation by the collision, and using (\ref{T-1}) and (\ref{H-const-2}),
\be
{\mathcal{H}}_{\mathrm{MR}}(\tau_c)= \mathcal{T}(\tau_c)=\frac{T_3}{g_s}\,\left(\frac{1}{\sqrt{1- \dot{\rho}(\tau_c)^2}}-1\right)=\frac{T_3}{g_s}\left(\frac{\pi}{g_s N \delta}-1\right). \label{H-rad-matter}
\ee
Because the collision occurs instantaneously, the scale factor after collision equals the scale factor before collision. In order to obtain a scale factor that accommodates the observable size of the universe starting with order Planck size, we need $\frac{\pi}{g_s N \delta} \gtrsim 10^{28}$.

\subsection{Stabilization of model universe\label{sec:stable}}

Now we use results in \cite{Hailu:2011kp} to argue that the model universe bounces and  stabilizes at a finite distance close to the IR boundary.
When the kinetic energy of the model universe is converted to matter and radiation by the collision, quarks on the D3-brane begin forming bound states of baryons with strings in-between and the wrapped D7-branes serving as sources of color.
Once baryons are formed, the size of the universe does not affect the size of the baryons; protons and neutrons continue to evolve in time with size of about a fermi. This is the scale of spatial separation $L$ between quarks in baryons. For given parameters $g_sN$ and $L$, the profile of strings between quarks and their closest distance to the D7-branes at the IR boundary is finite.
The model universe slightly bounces and stabilizes at a finite distance from the IR boundary at $\rho=\rho_e$  at time $\tau_e$ with $a(\tau_e)\approx a(\tau_c)=  \frac{\pi}{g_sN\, \delta}$ and  strong interactions (or the tension in the strings) overcoming the gravitational force from the background.

Let us briefly summarize the features we need from \cite{Hailu:2011kp}, with the focus here being a D3-brane located close to the IR boundary at $\rho=\rho_e <<\frac{2\pi}{g_sN}$ than at the UV boundary. See \cite{Hailu:2011kp} for more details. Consider a quark-antiquark pair located at $(x^1,\,x^2,\,x^3)=(x,0,0)=(\pm\frac{L}{2},0,0)$ on the D3-brane filling four-dimensional spacetime between the IR and UV boundaries with a string in-between. The string worldsheet action for a static configuration at some time $X^0\to\infty$ follows from the metric given by (\ref{metric10-1a}), and using Euclidean signature,
\be
S_2=\frac{1}{2\pi\alpha'}\int d\sigma^0d\sigma^1\, \sqrt{\det G_{MN}\partial_{a}X^M \partial_{b}X^N}=\frac{X^0}{2\pi}\int dx\, \sqrt{\cosh^2u+(\frac{d\rho}{dx})^2}\,,\label{S2-1}
\ee
where $(\sigma^0,\sigma^1)$ parameterize the string worldsheet and we choose $\sigma^0=x^0$ and $\sigma^1=x^1\equiv x$, $\rho=\rho(x)$, and use the identification $r_s=\sqrt{\alpha'}$ in \cite{Hailu:2011pn}. Minimizing the action (area bounded by a rectangular Wilson loop) using the Euler-Lagrange equation,
\be
\frac{dx}{d\rho}={\sech u \,\csch u}=\frac{2\pi}{g_sN}(\frac{1}{\rho}-(\frac{g_sN}{2\pi})^2\rho),\label{xrho-1}
\ee
\be
S_2=2\left(\frac{X^0}{2\pi}(x-\frac{1}{2}{\sech}^2 u)|_{x=0}^{L/2}\right)=\frac{X^0}{2\pi}\,L,\label{S2-2}
\ee
where we have taken $\rho_e<<\frac{2\pi}{g_sN}$ here (for the D3-brane near the IR boundary). Thus the potential energy of the quark-antiquark pair increases linearly with the spatial distance (or the string length) between them in four dimensions, $E=\frac{L}{2\pi}$. For contrast, $E=\frac{(L+1)}{2\pi}$ when the D3-brane is at the UV boundary, which is the same as the energy we have here for $L>>1$.

\begin{figure}[t]
\begin{center}
\includegraphics[width=5.50 in]{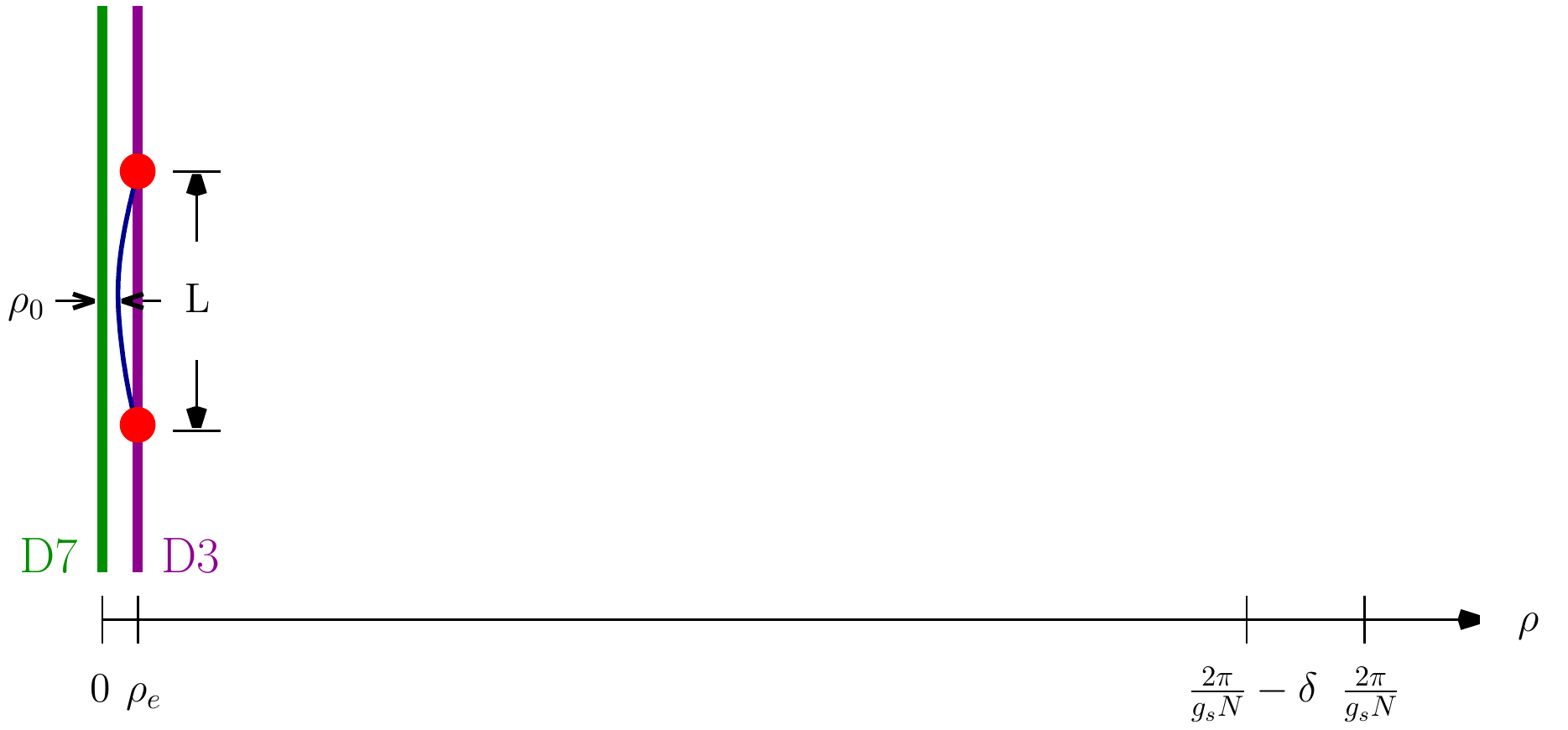}
\end{center}
\caption{Schematic description of the strings between quarks which stabilize the model D3-brane universe against gravitational force from the background of wrapped D7-branes. The D3-brane spontaneously appears in the UV region at $\rho=\frac{2\pi}{g_sN}-\delta$ and dynamically flows toward the IR and collides with the D7-branes at the IR boundary at $\rho=0$. The kinetic energy of the D3-brane is converted to matter and radiation by the collision and it stabilizes near the IR boundary at $\rho=\rho_e$ with the formation of baryons (a meson is shown here).}
\label{fig:stablized1}
\end{figure}

Solving (\ref{xrho-1}) with boundary condition that $x=\pm \frac{L}{2}$ at $\rho=\rho_e$, we write the static configuration of the string for the region $x\geq0$, it is symmetric about $x=0$, as
\be
x=\frac{2\pi}{g_sN}\left(\ln(\frac{\rho}{\rho_0})
-\frac{1}{2}(\frac{g_sN}{2\pi})^2(\rho^2-\rho_0^2)\right),\label{xrho-2}
\ee
where
\be
\frac{L}{2}= \frac{2\pi}{g_sN}\ln(\frac{\rho_e}{\rho_0})-\frac{1}{2}\frac{g_sN}{2\pi}\,
({\rho_e}^2-{\rho_0}^2)\label{Lrhoe0-1}
\ee
and $\rho_0$ measures how far the middle of the string is from the wrapped D7-branes.
This is exactly the same result as in \cite{Hailu:2011kp} for a D3-brane at the UV boundary which corresponds to $\rho_e=\frac{2\pi}{g_sN}$. For given finite values of $L$ and $g_sN$, the value of $\frac{\rho_e}{\rho_0}$ is finite too.
The string ends on the quark and the antiquark with slope $\frac{dx}{d\rho}|_{\rho=\rho_e}=\frac{2\pi}{g_sN}(\frac{1}{\rho_e}-(\frac{g_sN}{2\pi})^2\rho_e)$ which has magnitude of $\frac{2\pi}{g_sN}\frac{1}{\rho_e}$ for $\rho_e<<\frac{2\pi}{g_sN}$ we consider here and zero for $\rho_e=\frac{2\pi}{g_sN}$ and $\frac{dx}{d\rho}|_{\rho=0}=\infty$.
The strings ending on quarks on the D3-brane then serve as anchors which support the model universe against gravitational force from the background, recall that most of the mass of the visible matter in the universe is in the strong force (or in the strings in this picture). Quarks in baryons move on the D3-brane universe with mean separation that does not change as the universe evolves.
Schematic representation of a string in a quark-antiquark pair is shown in figure \ref{fig:stablized1} based on the above results.
\emph{Thus the model universe is stabilized against gravitational force from the background at a finite distance from the IR boundary by strong interaction which accounts for most of the visible mass of the universe.}

\subsection{The horizon, flatness, and inhomogeneity problems\label{sec:hf}}

Now we discuss how the model addresses the horizon, flatness, and inhomogeneity problems.
First let's briefly recall the nature of these problems in the hot big bang model and how the inflationary scenario addresses them. According to data from seven year Wilkinson Microwave Anisotropy Probe (WMAP) observations (combined with additional external data and fitting based on the standard cosmological model) \cite{Jarosik:2010iu}, the age of the universe is $13.75\pm 0.11$ billion years, the current total energy density is $\Omega_{\mathrm{tot}}=(1.0023^{+0.0056}_{-0.0054})\,\Omega_{c}$, the baryon density is $\Omega_{b}=(0.0456\pm0.0016)\,\Omega_{c}$, the dark energy density is $\Omega_{\Lambda}=(0.728^{+0.015}_{-0.016})\,\Omega_{c}$, the dark matter density is $\Omega_{dm}=(0.227\pm0.014)\,\Omega_{c}$, where $\Omega_{c}$ is the critical density which produces a flat universe in the Friedmann model, and the temperature power spectrum of the CMBR has a tilt parameterized by a scalar spectral index $n_s=0.963\pm0.012$. In addition, the size of the observable universe is about 46 billion light years and the CMBR is homogeneous and isotropic with mean temperature of about $2.7\,^{\circ}{\rm K}$ and relative variations of order $5\times 10^{-5}$.

In the hot big bang model the universe originated with sub-Planckian size and expanded to its current size. The horizon problem arises because if the current observable universe is evolved backward in time using the Friedmann equations for a radiation dominated case, one ends up with over $10^{28}$ causally disconnected Planck size regions which makes it hard to understand the observed homogeneity and isotropy. The flatness problem arises because the kinetic energy of the matter in the universe is related to the rate of expansion, and a flat universe with very near critical total energy density requires a highly fine-tuned initial velocity to an accuracy of over $10^{-54}\%$. The inhomogeneity problem is related to understanding the source of the small fluctuations of order $5\times 10^{-5}$.

The inflationary scenario addresses these problems by positing that the early universe underwent a period of accelerating expansion (inflation) which stretches its initially order Planck size by a factor of over $10^{28}$. The inflation arises due to a scalar field $\phi$ rolling on a potential energy density $\mathcal{V}(\phi)$ that produces negative pressure and the sources of inhomogeneities are quantum fluctuations which get stretched during inflation and early phase of Friedmann evolution. There are important requirements that the potential needs to satisfy in the standard slow-roll inflationary scenario in order to address the horizon, flatness, and inhomogeneity problems, see \cite{Liddle:2000cg} for instance,
\be
\epsilon \equiv \frac{M_4^2}{2}\left(\frac{\mathcal{V}'}{\mathcal{V}}\right)^2 <<1,\label{slowroll-ep-1}
\ee
\be
|\eta| \equiv |{M_4^2}\frac{\mathcal{V}''}{\mathcal{V}}| <<1,\label{slowroll-eta-1}
\ee
\be
\delta_{H}\equiv\frac{1}{\sqrt{75}\,\pi}\frac{1}{M_4^3}\frac{\mathcal{V}^{3/2}}
{\mathcal{V}'}\sim 5\times 10^{-5},\label{slowroll-Hd-1}
\ee
where $M_4$ is the Planck scale in four dimensions expressed in units of $r_s^{-1}={\alpha'}^{-1/2}$ here and a prime denotes differentiation with respect to $\phi$. In standard inflationary models, the potential needs to be flat enough for a wide enough range of $\phi$ in order to stretch the scale factor by over $10^{28}$ and, at the same time, it needs to vary slowly enough at the time when quantum fluctuations exit the Hubble radius in order to produce the desired level of inhomogeneities. The observed spectral tilt in the CMBR spectrum puts the constraint
\be
n_s-1=2\eta-6\epsilon\approx-0.04.\label{ns-1aa}
\ee

In our case, first we show that in the region where accelerating expansion of the model universe occurs, we can use the methods of slow-roll inflation. Let us see how long the accelerating expansion lasts using the equation of motion given by (\ref{rho-eqn-1}), since $\ddot{\rho}=0$ at the transition from accelerating to decelerating expansion, which gives
\be
\dot{\rho}(\tau_{eae})=-\sqrt{\frac{2}{3}}\,\cosh u\,,\label{drho-1}
\ee
where $\tau_{eae}$ is the time at the end of the accelerating expansion phase. We have $0\le \dot{\rho}^2\sech^2 u\le \frac{2}{3}$ during this time. Therefore, we can make a Taylor expansion of the action to leading order in $\dot{\rho}^2\sech^2 u$ in this region and (\ref{DBI-D3-1a}) reduces to
\be
S=\int  d^4x\,\left(\frac{1}{2}\frac{T_3}{g_s}\dot{\rho}^2-\mathcal{V} \right),\label{DBI-D3-1aTaylor}
\ee
where $\mathcal{V}$ is given by (\ref{V-1}). Therefore, we can use the methods of slow-roll inflation during the accelerating expansion phase where we identify the scalar field
\be
\phi= \sqrt{\frac{T_3}{g_s}}\,{\rho}\,,\label{phirho-1}
\ee
in units of $r_s^{-1}$ (recall that $T_3$ and $x^\mu$ are dimensionless and measured in units of $r_s^{-4}$ and $r_s$ respectively in our notation; we have $\phi= \sqrt{\frac{T_3}{g_s}}\,{\rho}\,r_s$ with $\rho = 3 \ln(\frac{r}{r_s})$ when expressed in terms of dimenionful variables). The potential given by (\ref{V-1}) can rewritten in terms of $\phi$ as
\be
\mathcal{V}=\frac{T_3}{g_s}\,\frac{1}{1-\frac{g_s}{T_3}(\frac{g_sN}{2\pi}\phi)^2}\,.\label{V-1a}
\ee

First, with appropriate choice of parameters such that $\frac{\pi}{g_sN\,\delta}\gtrsim 10^{28}$, the scale factor grows by a factor of over $10^{28}$ before a big bang collision followed by formation of baryons and stabilization near the IR boundary where a hot universe begins Friedmann evolution with a scale factor $a(\tau_c)=\frac{\pi}{g_sN\,\delta}\,a(\tau_i)$. \emph{Thus the model addresses the horizon problem.}

Moreover, because the background preserves $\mathcal{N}=1$ supersymmetry in four dimensions by a balance between fluxes and torsion, both the model D3-brane universe and the wrapped D7-branes stay flat along $\mathds{R}^{3}$, the model universe expands from order Planck scale size by a large factor, and the production of matter and radiation due to collision of the flat branes also occurs homogeneously except for fluctuations. \emph{Thus the model addresses the flatness problem.}
In addition, the speed $\dot{\rho}(\tau_c)$ (and the kinetic energy and rate of expansion) at the time of collision is insensitive to the initial speed $\dot{\rho}(\tau_i)$. That is because $\sech u(\tau_i)\approx 0$ and the value of $\dot{\rho}(\tau_i)^2 \sech^2 u(\tau_i)<<1$ insensitive to the value of $\dot{\rho}(\tau_i)$; the kinetic energy of the model universe depends on the parameter $\delta$ which describes how close its initial position is to the UV boundary for a given $g_sN$.

Next we focus on the generation of inhomogeneities that serve as sources of thermal fluctuations in the CMBR and seeds for large scale structure formation.
For $\frac{g_sN\,\delta}{2\pi}<<1$, the acceleration at the beginning is given by (\ref{ddrho-dda-1}).
With (\ref{ddrho-dda-1}) and (\ref{drho-1}), we can estimate the order of magnitude for the duration of the flow in which the expansion is accelerating with $\ddot{a}>0$,
$\tau_{eae}-\tau_i \sim \sqrt{\frac{2\,gsN\,\delta^3}{3\pi}}$.
We also estimate the order of magnitude of how much the scale factor grows during this time and the rate at which it is changing at  $\tau_{eae}$ using (\ref{drho-1}) in (\ref{a-1}) and (\ref{adot-analy-1}),
$\frac{a(\tau_{eae})}{a(\tau_{i})}\sim \sqrt{3}$ and $\dot{a}(\tau_{eae}) \sim \sqrt{\frac{2\pi}{3\,gsN\,\delta^3}}$.
Therefore, the scale factor does not grow very much during the phase of accelerating expansion, its enormous growth occurs in the decelerating expansion phase where $\dot{a}$ starts with a large value for appropriately small $g_sN\,\delta^3$ and stays large for much of the flow until the model universe gets close to the IR boundary. This is different from standard inflationary scenarios where the enormous growth of the scale factor is produced during exponentially accelerating expansion.

\begin{figure}[t]
\begin{center}
\includegraphics[width=5.5 in]{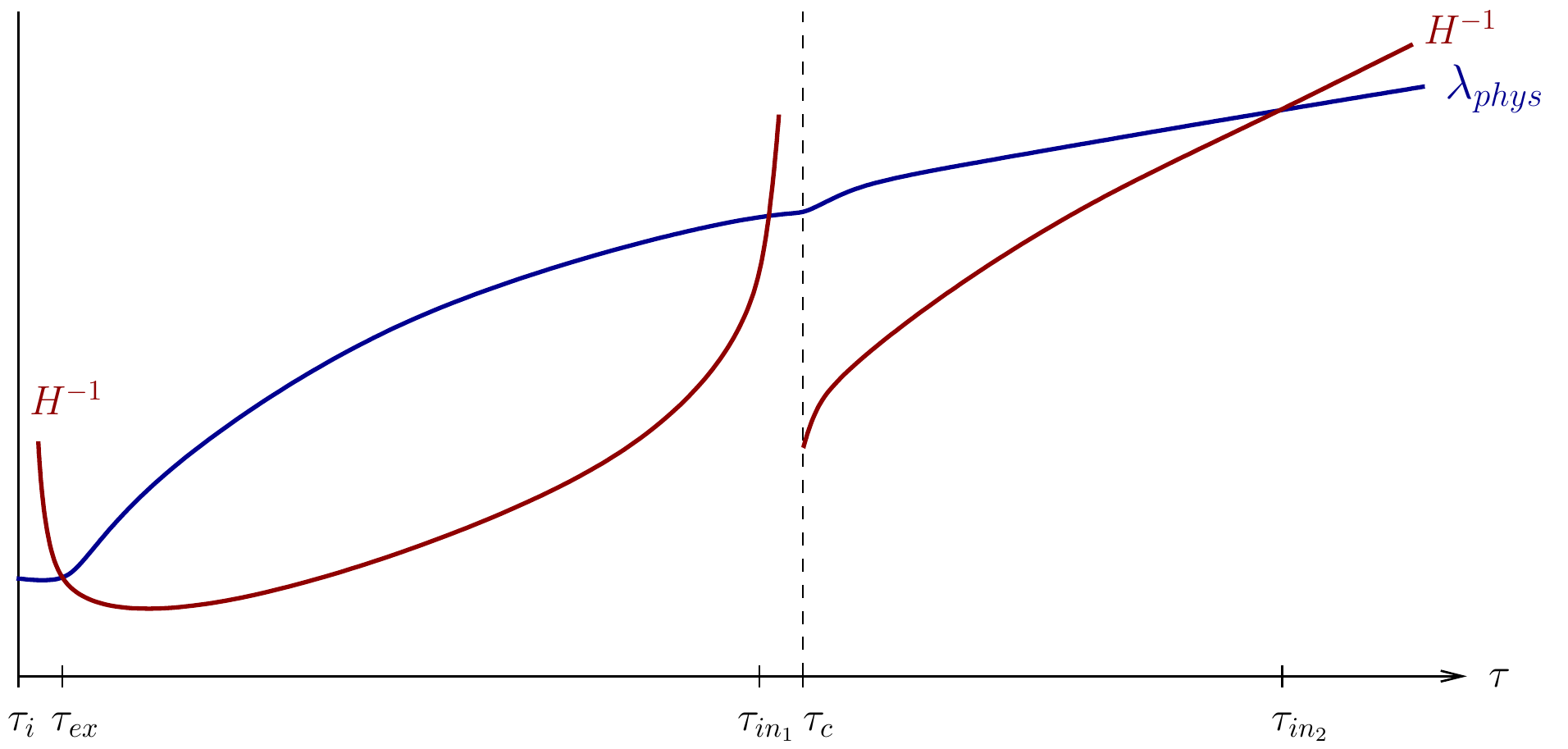}
\end{center}
\caption{Schematic description of how fluctuations exit and enter the Hubble radius. Quantum fluctuations with physical wavelength $\lambda_{phys}$ of order Planck size cross the Hubble radius $H^{-1}$ at time $\tau_{ex}$ close to $\tau_i$ during the accelerating expansion phase as the Hubble radius decreases. The Hubble radius increases as the model universe approaches the IR boundary and the fluctuations enter the curvature at time $\tau_{in_{1}}$  for a brief time during which the scale factor barely changes just before collision (big bang). The kinetic energy of the D3-brane is converted to matter and radiation by the collision at time $\tau_c$, the rate of expansion jumps from zero to a finite value while the physical wavelength of fluctuations $\lambda_{phys}\sim a$ is continuous, and the fluctuations exit the Hubble radius again. The Hubble radius then increases and the fluctuations reenter the curvature at a later time $\tau_{in_{2}}$ during Friedmann evolution.}
\label{fig:densitypert1}
\end{figure}

Quantum fluctuations that serve as sources of thermal fluctuations in the CMBR and seeds for large scale structure formation are produced during the accelerating expansion phase. In this phase the Hubble radius $H^{-1}$ decreases while the scale factor $a$ stays nearly constant.
Therefore, quantum fluctuations of order Planck size exit the Hubble radius at some time $\tau$, where $\tau_i<\tau<\tau_{eae}$ when their physical wavelength $\lambda_{phys}\sim H^{-1}$.  In order for the fluctuations to get stretched and grow with the scale factor, they need to stay outside the Hubble radius. This occurs if $H^{-1}=\frac{a}{\dot{a}}<\lambda_{phys}\sim a$. Because both the fluctuations and the model universe begin with order Planck size, we need $\frac{1}{\dot{a}}<1$ which using (\ref{adot-analy-1}) becomes $\frac{\delta}{\tanh u \,|\dot{\rho}|}<1$. Notice that $\frac{\delta}{\tanh u \,|\dot{\rho}|}$ starts with infinitely large value at time $\tau_i$ and falls during the accelerating expansion phase as $|\dot{\rho}|$ increases, and because $\tanh u\approx 1$ in the UV, we require $|\dot{\rho}(\tau_{eae})|>\delta$ so that fluctuation exit the Hubble radius. Furthermore, we want $\delta<<1$ (and $\delta<1$ is good enough) so that the fluctuations stay outside the Hubble radius during most of the expansion as the model universe gets close to the IR boundary where $|\dot{\rho}|\approx1$, $\dot{a}\to 0$ and $\tanh u \to 0$. In this case, the fluctuations stay outside the Hubble radius for almost all of the decelerating expansion phase, since $\dot{a}>>1$ during most of the flow until collision. Close to the IR boundary, $\tanh u\to0$ and the fluctuations enter the Hubble radius. The collision converts the kinetic energy of the D3-brane due to its motion along the radial direction to matter and radiation moving in four dimensions confined to the model universe which produces a discontinuity in $\dot{a}$ while $a$ stays continuous, and the fluctuation exit the Hubble radius again. Furthermore, because fluctuations with larger $\lambda_{phys}$ exit the Hubble radius earlier during the accelerating expansion phase and have larger amplitude when they reenter, the power spectrum has red tilt consistent with observations.
A schematic plot which shows how the Hubble radius $H^{-1}$ and the physical wavelength of fluctuations $\lambda_{phys}$ evolve in time is shown in figure \ref{fig:densitypert1}. \emph{Thus the model addresses the inhomogeneity problem.}

Because the background geometry is compact, $M_4$ can be expressed in terms of the volume $V_6$ from the extra space and  we have a finite Newton's constant in four dimensions. Using the metric given by (\ref{metric10-1a}), the ten-dimensional volume is
\be
V_{10}=V_4\,  (2\pi)^5r_s^6\,\int_{0}^{\frac{2\pi}{g_sN}} \sqrt{1-(\frac{g_sN}{2\pi}\,\rho)^2}\,d\rho=V_4\,V_6\,,\qquad V_6 ={4\pi^5\, g_sN}\,,\label{V10-6-4-a}
\ee
where $V_4=\int d^4x\,\sqrt{-g}$ is the volume of the 4D spacetime, with $V_4$ and $V_6$ measured respectively in units of $r_s^4$ and $r_s^6$, and we use the mapping $\alpha' = r_s^2$ in \cite{Hailu:2011pn}. We then have
\be
{M_4^2}=\frac{2V_6}{(2\pi)^7 g_s^2}=\frac{ g_sN}{16\pi^2 g_s^2\,}\,,\label{M4-1}
\ee
with $M_4$ measured in units of $r_s^{-1}$.
\emph{Thus the background gives a finite value of Newton's gravitational constant in four dimensions which reproduces the experimentally measured value with appropriate choice of parameters.}
A smaller value of $g_sN$ ('t Hooft coupling at the UV boundary) corresponds to smaller curvature in which the supergravity description is valid on the whole region of interest away from the UV boundary.

We can obtain the total time the model universe takes from time $\tau_i$ to the beginning of Friedmann evolution using the equation of motion. Because the range of $\rho$ it moves across is $\frac{2\pi}{g_sN}-\delta\sim\frac{2\pi}{g_sN}$ and the speed $\dot{\rho}\sim-1$ for most of the flow to leading order in $\frac{g_sN\,\delta}{2\pi}$, the order of magnitude of the total time is
\be
\tau_{c}-\tau_{i}\sim\frac{2\pi}{g_sN}.\label{Dx0ie-1}
\ee

Using the relation between $u$ and $\rho$ given by (\ref{u-rho-1}) and the definition of $\phi$ given by (\ref{phirho-1}),
\be
u'=  \frac{g_sN}{2\pi}\sqrt{\frac{g_s}{T_3}}\,\cosh^2u\,,
\ee
where again a prime denotes differentiation with respect to $\phi$,
and with the potential in (\ref{V-1}) we have
\be
\mathcal{V}'=-\frac{g_sN}{\pi}\sqrt{\frac{T_3}{g_s}}\,\cosh^3 u \sinh u,\label{Vp-pp-1a}
\ee
\be
\mathcal{V}''=-\frac{(g_sN)^2}{2\pi^2}\,(2\cosh2u-1)\cosh^4 u\,,\label{Vp-pp-1b}
\ee
\be
\frac{\mathcal{V}'}{\mathcal{V}}=-\frac{g_sN}{2\pi}\sqrt{\frac{g_s}{T_3}}\, \sinh 2u\,,\qquad
\frac{\mathcal{V}''}{\mathcal{V}}=-\frac{(g_sN)^2}{2\pi^2}\frac{g_s}{T_3}\,(2\cosh2u-1)\cosh^2 u\,,\label{Vp-pp-2}
\ee
where the minus sign is because the model universe is flowing toward the IR along which the potential decreases.

The valid parameters space satisfying the constraints given by (\ref{slowroll-ep-1})-(\ref{ns-1aa}), (\ref{M4-1}),  and the level of expansion can be investigated numerically.
Our interest here is to find analytic expressions to leading order in $\frac{g_sN\,\delta}{2\pi}$ in the region the model universe undergoes accelerating expansion and quantum fluctuations exit the Hubble radius. This occurs near $\rho=\frac{2\pi}{g_s N}-\delta$ where, to leading order in $\frac{g_sN\,\delta}{2\pi}$,

\be
\mathcal{V}=\frac{T_3}{g_s}\,\frac{\pi}{g_s N \delta}\,,\quad
\mathcal{V}'=-\sqrt{\frac{T_3}{g_s}}\,\frac{\pi}{g_s N \delta^2}\,,\quad
\mathcal{V}''=-\frac{2\pi}{g_s N \delta^3}\,,\label{V-Vp-pp-2a}
\ee
\be
\frac{\mathcal{V}'}{\mathcal{V}}=-\sqrt{\frac{g_s}{T_3}}\,\frac{1}{\delta}\,,\quad
\frac{\mathcal{V}''}{\mathcal{V}}=-{\frac{g_s}{T_3}}\,\frac{2}{\delta^2}
\,.\label{V-Vp-pp-2b}
\ee

With (\ref{V-Vp-pp-2a}) and (\ref{V-Vp-pp-2b}), the constraints (\ref{slowroll-ep-1})-(\ref{ns-1aa}) together with (\ref{M4-1}) become, putting $r_s$ back which we do from now on with $x^\mu$ containing $r_s$ and the metric (\ref{metric10-1a}) rewritten as in \cite{Hailu:2011pn}, $ds_{10}^2={\cosh u}\,g_{\mu \nu}dx^\mu dx^\nu
+{r_s^2}\, \sech u\,\left({d\rho^2}+d\psi^2+d\varphi_1^2+d\varphi_2^2
+d\varphi_3^2+d\varphi_4^2\right)$,
\be
\epsilon= \frac{M_4^2}{2\,\delta^2}\,{\frac{g_s}{T_3 r_s^2}}<<0\,,\label{slowroll-ep-1a}
\ee
\be
|\eta|=\frac{2M_4^2}{\delta^2}\,{\frac{g_s}{T_3 r_s^2}}=4\,\epsilon<<0\,,\label{slowroll-eta-1a}
\ee
\be\delta_H= \frac{1}{\sqrt{75\,\pi}M_4^3}\,\sqrt{\frac{\delta}{g_sN}}\,
{\frac{T_3 r_s}{g_s}}\sim\,5\times 10^{-5}\,,\label{slowroll-Hd-1a}
\ee
\be
{M_4^2}=\frac{ g_sN}{16\pi^2r_s^2 g_s^2\,}\,.\label{M4-1a}
\ee
The model has the following parameters: $g_sN$, $\delta$, $T_3$, $r_s$, and $N$ (or $g_s$). We will consider an example of numerical values of these parameters and test the model in section \ref{sec:numericalparm}.

\subsection{Dark energy and Friedmann evolution\label{sec:darke}}

Once the model universe stabilizes, it follows Friedmann evolution with a cosmological constant term coming from the remaining potential energy density of the D3-brane in the background which we interpret as dark energy.
When it reaches the IR boundary almost all of the initial potential energy density is converted to kinetic energy density and the remaining potential energy density can be read off from the Hamiltonian given by (\ref{V-1}) with $\rho<<\frac{2\pi}{g_sN}$,
\be
\Lambda=\frac{T_3}{g_s}=\frac{a(\tau_i)}{a(\tau_c)}\,\mathcal{H}(\tau_i).\label{Lam-1a}
\ee
\emph{Thus the potential energy density of the model universe gives dark energy density that is smaller than the initial potential (or total) energy density by a ratio of the scale factor at the beginning near the UV boundary to that at the time of collision at the IR boundary.}

We then use this in the Friedmann equations for a flat universe,
\be
H^2=\frac{1}{3\,M_4^2}\,(\varepsilon_M+{\Lambda}),\label{Friedmann-1}
\ee
\be
\frac{\ddot{a}}{a}=-\frac{1}{6\,M_4^2}\,(\varepsilon_M+3p-2{\Lambda}),\label{Friedmann-2}
\ee
where $H\equiv\frac{\dot{a}}{a}$ is the Hubble constant, $\varepsilon_M$ is matter density, and $p$ is pressure.

For time $\tau>\tau_e\approx \tau_c$, the matter-radiation density falls with increasing spatial volume of the model universe while the dark energy density stays constant. From (\ref{H-rad-matter}),  (\ref{Lam-1a}), and (\ref{a-rho-ts-1}) we see that
\be
\frac{\mathcal{H}_{MR}(\tau_{c})}{\Lambda}=\frac{\pi}{g_s N \delta}-1= a(\tau_{c})-1\label{HmroverLam-1}
\ee
Because the total amount of matter-radiation for time $\tau>\tau_{c}$ should stay constant and $a(\tau)>>1$ (in our notation) during Friedmann evolution, the ratio of the matter-radiation energy density  now at time $\tau_n$ to that at the beginning of Friedmann evolution is
\be
\frac{\mathcal{H}_{MR}(\tau_n)}{\mathcal{H}_{MR}(\tau_{c})}=\left(\frac{a(\tau_{c})}{a(\tau_n)}\right)^3,\label{HmroverLam-2}
\ee
On the other hand, cosmological observations show that the current ratio $\frac{\mathcal{H}_{MR}(\tau_n)}{\Lambda}\approx \frac{3}{7}$, where we have included cold dark matter in $\mathcal{H}_{MR}$. Thus the scale factor today is related to the scale factor at the beginning of Friedmann evolution,
\be
a(\tau_n)\approx \left(\frac{3}{7}\right)^{1/3}a(\tau_{c})^{4/3}=\left(\frac{3}{7}\right)^{1/3}
\left( \frac{\pi}{g_sN\, \delta}\right)^{4/3}.
\ee

According to the picture here, a constant dark energy density existed since collision (big bang), and its contribution to the evolution of the universe becomes proportionally more significant as the universe expands and the matter-radiation density become smaller.
\emph{Thus the remaining potential energy density serves as a source of dark energy that produces accelerating expansion of the model universe during dark energy dominated phase.}

\subsection{Summary of constraints from observations\label{sec:sumconst}}

First let us recall the parameters of the model: $g_sN$, $\delta$, $T_3$, $r_s$, and $N$. The 't Hooft coupling at the UV boundary is $g_sN$ and its magnitude determines the range of $\rho$, $0\le \rho\le\frac{2\pi}{g_sN}$. A smaller value of $g_sN$ corresponds to a larger range and a smaller curvature.  The radius of the internal space at the IR boundary is $r_s$. $N$ is the number of wrapped D7-branes at the IR boundary. The model universe spontaneously appears in the UV region at $\rho=\frac{2\pi}{g_sN}-\delta$ at time $\tau_i$ with $\delta<<\frac{2\pi}{g_sN}$. The initial energy density of the D3-brane is all potential and $\mathcal{H}(\tau_i)=\frac{T_3}{g_s}\,\frac{\pi}{g_s N \delta}$, and its magnitude is larger than the dark energy density $\Lambda=\frac{T_3}{g_s}$ during Friedmann evolution by a factor of $\frac{\pi}{g_s N \delta}$.

Let us summarize the constraints that follow from the desired level of expansion needed to address the horizon and flatness problems for a universe that begins with Planck size, cosmological data on inhomogeneities, the measured value of Planck scale (or Newton's constant) in four dimensions, and observed level of current accelerating expansion,
\be
\frac{a(\tau_c)}{a(\tau_i)}=\frac{\pi}{g_sN\,\delta}\gtrsim 10^{28}\,,\label{a-1aa}
\ee
\be
\epsilon=\frac{M_4^2}{2\,\delta^2}\,{\frac{g_s}{T_3 r_s^2}}
=\frac{1}{4}|\eta|\,,\label{ep-1aa}
\ee
\be
\delta_H= \frac{1}{\sqrt{75\,\pi}M_4^3}\,\sqrt{\frac{\delta}{g_sN}}\,
{\frac{T_3 r_s}{g_s}}\sim 5\times 10^{-5}\,,\label{dh-1aa}
\ee
\be
{M_4}=\sqrt{\frac{g_sN}{16\pi^2g_s^2r_s^2}}=\frac{1}{1.6\times 10^{-35}\,m}= 1.2\times10^{19}\,GeV\,,\label{M4-1aa}
\ee
\be
{\Lambda}=\frac{T_3}{g_s}\sim 10^{-120}\,M_4^4\,.\label{Lam-1aa}
\ee
The constraints on the values of $\epsilon$ and $\eta$ gives the spectral tilt in the CMBR,
\be
n_s-1=2\eta-6\epsilon=-14\,\epsilon\approx-0.04.\label{ns-1}
\ee
In addition, requiring that the fluctuations exit the Hubble radius during most of the expansion as shown in figure \ref{fig:densitypert1} requires
\be
\delta<1\,.\label{delta-const-1}
\ee

\subsection{An example of numerical values of parameters\label{sec:numericalparm}}

The model has already been constructed in previous sections. Here we discuss one example of numerical values of the parameters and test the model.

Consider the following choice of parameters,
\ba
&g_sN=7.2\times 10^{-28}\,,\qquad T_3=5.9\times 10^7\,r_s^{-4}\,,\qquad  M_4= 3.0\times 10^{19}\,r_s^{-1}\,,&\nn\\ &\delta=4.3\times 10^{-1}\,,\qquad N=10^7,\,\qquad r_s=4.7\times 10^{-16}\,m.&
\ea
First, notice that $r_s$ (the radius of the internal space at the IR boundary) is about the size of baryons which is a reasonable value for identifying $r_s$ with the nonperturbative scale of the gauge theory whereby strong interaction stabilizes the model universe near the IR boundary against gravitational force from the background. 
The volume of the extra space seen by a four-dimensional observer and given in (\ref{V10-6-4-a}) is $V_6 = {4\pi^5r_s^6\, g_sN}=(4.6\times10^{-20}\,m)^6$.  We also have the measured numerical value of Planck scale (or Newton's gravitational constant) in four-dimensions ${M_4}=\sqrt{\frac{ g_sN}{16\pi^2r_s^2 g_s^2}}=3.0\times 10^{19}\,r_s^{-1}= \frac{1}{1.6\times 10^{-35}\,m}= 1.2\times10^{19}\,GeV$ and a desirable level of expansion $\frac{a(\tau_c)}{a(\tau_i)}=\frac{\pi}{g_sN\,\delta}=10^{28}$. It takes a time of about $\frac{2\pi}{g_sN}\frac{r_s}{c}=3.8$ hours for the model universe to flow from  near the UV boundary to collision at the IR boundary which is taken as a big bang and the beginning of a hot universe with matter and radiation.
The slow-roll constraints on the potentials given by (\ref{slowroll-ep-1a})-(\ref{slowroll-Hd-1a}) are satisfied,  $\epsilon=\frac{M_4^2}{2\,\delta^2}\,{\frac{g_s}{T_3 r_s^2}}=2.9\times 10^{-3} <<1$ and $|\eta|=\frac{2M_4^2}{\delta^2}\,{\frac{g_s}{T_3 r_s^2}}=4\epsilon=1.1\times 10^{-2} <<1$. For the spectral tilt in the CMBR,
$n_s-1=2\eta-6\epsilon=-0.04$ and for the density perturbation,
$\delta_H=5\times 10^{-5}$. Moreover, because $\delta=0.43<1$, quantum fluctuations which exit the Hubble radius during the accelerating expansion phase stay outside and grow with the scale factor as schematically shown in figure \ref{fig:densitypert1}.
Thus the above set of parameters produces the desired level of expansion for a universe that begins with Planck size in addition to the observed experimental values of scalar density perturbation, spectral tilt, and Newton's gravitational constant in four dimensions.

The dark energy density during Friedmann evolution is
$\mathcal{H}_{\mathrm{DE}}=\frac{T_3}{g_s}=8.1\times 10^{41}\, r_s^{-4}= 1.1\times 10^{-36}\, M_4^4$.
The total energy density, which stays constant during the flow until the beginning of Friedmann evolution and equals the potential energy density at time $\tau_i$, is
$
\mathcal{H}(\tau)=\frac{T_3}{g_s}\,\frac{\pi}{g_s N \delta}= 1.1\times 10^{-8}\, M_4^4=(0.01\,M_4)^4$
for $\tau_i\le\tau_c$.
Thus, for the numbers we have here, the model universe flows with order Planck total energy density which is initially all potential and almost all of it becomes kinetic at the IR boundary.
The kinetic energy is then converted to matter and radiation by the collision. The remaining potential energy density becomes dark energy density that is smaller than $\mathcal{H}(\tau)$ by a factor of $10^{-28}$; i.e., $\Lambda=1.1\times 10^{-36}\, M_4^4$, which is reduced in the right direction but not small enough to accommodate the observed value of order $10^{-120}\,M_4^4$. Recall that $\mathcal{\mathcal{H}_{\mathrm{DE}}}=\Lambda=\frac{a(\tau_i)}{a(\tau_c)}\mathcal{H}(\tau)$ and a smaller value of dark energy density can be obtained by taking smaller $g_sN\,\delta$. One needs $\frac{a(\tau_c)}{a(\tau_i)}\sim 10^{120}$ with $\mathcal{H}(\tau)$ of order Planck energy density in order to address the dark energy problem, and this needs to be done while the other constraints on the desired level of expansion, density perturbations, and spectral tilt are satisfied.

Finally, because $g_sN$ (the 't Hooft coupling at the UV boundary) is small, the curvature throughout the region in which the model universe flows is small and $\alpha'$ corrections are negligible. The string coupling at the UV boundary is $g_s=\frac{g_sN}{N}=7.2\times 10^{-35}$, which is small. Neither the gauge theory nor the gravity theory predicts the absolute magnitude of the string or gauge coupling (the renormalization group flow relates the coupling at one scale to the coupling at another) and we have picked a value that is suitable for our discussion here. The string we have here is that of strong interactions with tension of $\frac{1}{2\pi\alpha'}=\frac{1}{2\pi r_s^2}$, where $r_s$ is given above. Furthermore the D3-brane tension $T_3=(\frac{1}{0.01\,r_s})^{4}$, which has the same order of mass scale. When the D3-brane appears in the supergravity background, its energy density becomes order Planck scale, as we saw above, due to gravitational interactions. The four-dimensional gauge coupling runs and both the gauge and 't Hooft couplings of the gauge theory living on the wrapped D7-branes become $\mathcal{O}(1)$ and larger close enough to the IR boundary as $r\to r_s$ or $\rho\to 0$, see \cite{Hailu:2011pn} for details. Moreover, because $N>>1$ both string loop corrections and backreaction from introducing a single D3-brane are negligible. We would also like to point out that lattice QCD computations for physical observables such as mass ratios of glueballs in the $N\to \infty$ limit are not significantly different from the case of $N=3$, see \cite{Lucini:2010nv} for instance.

\section{Conclusions\label{sec:concl}}

The fact that the evolution of the universe cannot be accounted by its visible matter and energy content might partly be due to forces external to it and
the supergravity background argued to be dual to pure $\mathcal{N}=1$ $SU(N)$ gauge theory in \cite{Hailu:2011pn,Hailu:2011kp,Hailu:2012jc} produces accelerating expansion followed by smooth transition to decelerating expansion on a D3-brane model universe near the UV boundary as it flows toward the IR with the dynamics dictated by the background geometry through the DBI action without introduction of additional terms by hand. The model presented addresses the horizon, flatness, and inhomogeneity problems and provides a possible explanation to dark energy.

The model universe flows to the IR boundary with constant total energy density with
almost all of the initial potential energy density becoming kinetic energy density. The kinetic energy is then converted to matter and radiation by collision (big bang) with the wrapped D7-branes at the IR boundary. This conversion of kinetic energy to matter and radiation is one of the assumptions we have made in constructing the model, and it is analogous to the process of reheating which is similarly an assumption in standard inflationary models.
The remaining potential energy density gives a cosmological constant which is interpreted as dark energy density.
The strong interaction which accounts for most of the visible mass of the universe is modeled by strings between quarks in baryons (whose size does not change as the universe expands) which stabilizes the model universe at a finite distance from the IR boundary against gravitational force from the background. The model universe then follows Friedmann evolution.
Because the dark energy density stays constant while the matter-radiation density falls during Friedmann expansion, the contribution of dark energy to the cosmic  evolution increases with increasing time and produces accelerating expansion in dark energy dominated phase. The ratio of dark energy density during Friedmann evolution to the initial energy density when the model universe appears near the UV boundary (or during the flow) is equal to the ratio of the scale factor initially to the scale factor at the beginning of Friedmann evolution.

Cosmological observations from the desired level of expansion for a universe that begins with order Planck size and experimental values of density perturbations, spectral tilt, dark energy, and Newton's gravitational constant in four dimensions put constraints on the parameters of the model. We have written down the constraints using results from slow-roll inflation and discussed one example of numerical values of parameters which accommodates all the constraints except that the dark energy density is not small enough (but reduced from Planck scale in the right direction). Smaller value of dark energy density can be obtained by taking smaller $g_sN\,\delta$. Considering the importance of addressing the problems discussed here simultaneously in the same setting, in addition to providing a dual theory for studying the nonperturbative dynamics of $\mathcal{N}=1$ supersymmetric $SU(N)$ gauge theory,
it will be interesting to investigate the model, its parameters space, and available cosmological data further.

Because the total energy of a Planck size model universe is comparatively much smaller than the energy content of a large number of wrapped D7-branes, it is plausible that the model universe might originate from quantum fluctuations of the background. The spontaneous appearance of a D3-brane in the background is the other assumption we made in constructing the model; in the standard or other cosmological models, the origin or existence of a universe is always an assumption.

The dynamics discussed in section \ref{sec:eqnofm} might also be useful to investigate possibilities for cyclic universe scenarios such as implementing some features of the ekpyrotic scenario \cite{Khoury:2001wf} and its cyclic extension without requiring a singular beginning \cite{Steinhardt:2001st}.  The size between the IR and UV boundaries and the time interval between collisions can be made large by taking $g_sN$ to be appropriately small. If only part of the kinetic energy of the D3-brane is converted to matter and radiation by the collision, then the D3-brane would continue bouncing back and forth (returning from a point which gets further away from the UV boundary after each bounce). See appendix \ref{app:cyc} for more.

Finally, the results in this paper together with \cite{Hailu:2011pn,Hailu:2011kp,Hailu:2012jc} lead us to a  picture of D3-brane model universe with matter and radiation on it, the wrapped D7-branes serving as sources of color to strong interactions, and dark energy interpreted as potential energy of the model universe in the background. Thus the supergravity background can be used to investigate both the nonperturbative dynamics of strong interactions and cosmology simultaneously in the same setting.

\newpage

\appendix

\section{Cyclic universe\label{app:cyc}}

\begin{figure}[t]
\centering
\subfloat{\label{fig:rho3nn}\includegraphics[width=0.6\textwidth]
{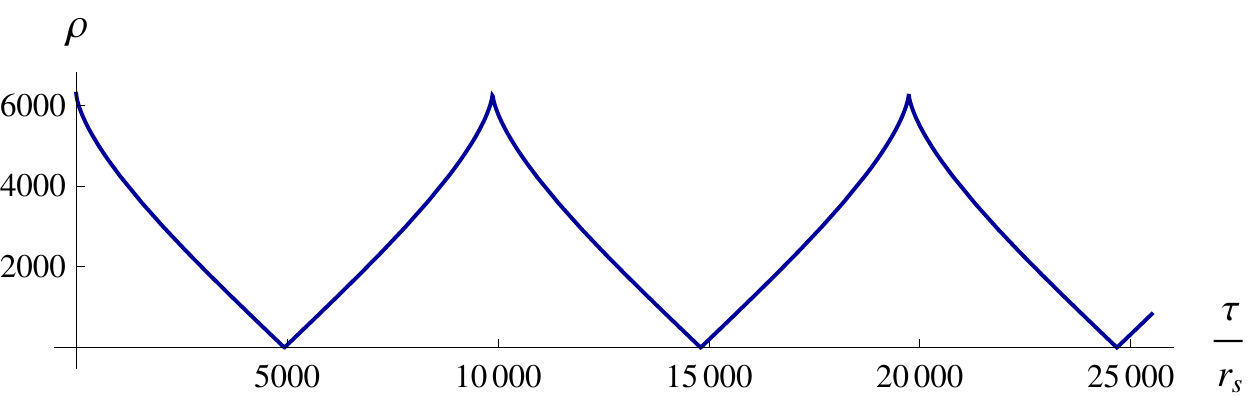}}
\\
\subfloat{\label{fig:rho3dsnn}\includegraphics[width=0.6\textwidth]
{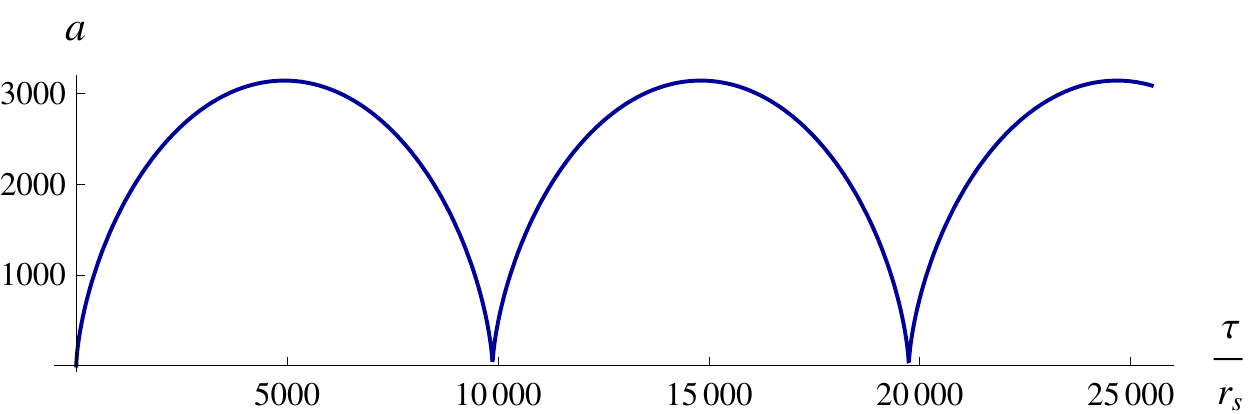}}
\caption{Cyclic evolution of $\rho$ and $a$ for $g_sN=10^{-3}$ and $\delta=1$.}\label{figs:cycgsp1dm3}
\end{figure}

Here we discuss how the dynamics discussed in section \ref{sec:eqnofm} might also be useful to construct a dynamical cyclic universe.
Consider a D3-brane that is located near the UV boundary at $\rho(\tau_i)=\frac{2\pi}{g_sN}-\delta$ at time $\tau_i$ with $\dot{\rho}(\tau_i)=0$, where $\delta<<\frac{2\pi}{g_sN}$, $\tau_i$ does not need to be a beginning. The D3-brane flows toward the IR with the scale factor of the three-dimensional space given by (\ref{a-1}) increasing to $a(\tau_c)=\frac{\pi}{g_sN\,\delta}\,a(\tau_i)$ at time of collision at the IR boundary. Part of the potential energy density of the D3-brane is converted to kinetic energy density when the D3-brane reaches the IR boundary. Considering a case in which only part of the kinetic energy is converted to matter and radiation by the collision, the D3-brane bounces back with a reduced radial speed and comes to a stop at a further point from the UV boundary than its location at time $\tau_i$ and then begins to flow back to the IR again. The cycle repeats. Thus the background provides a setting for cyclic universe scenarios.
This might, for instance, be useful to
investigate possibilities for implementing some features of
the ekpyrotic model \cite{Khoury:2001wf} and its cyclic extension \cite{Steinhardt:2001st}. Here, the D3-brane bouncing back and forth might be taken as the universe and
its cosmic evolution is dynamically dictated by the warped background geometry of the wrapped D7-branes. The size between the IR and UV boundaries and the time interval between collisions can be made large by taking $g_sN$ to be appropriately small. The warping of space is such that the volume of the extra space seen by a four-dimensional observer on the D3-brane and the curvature of the background decease with increasing distance between the two boundaries (or decreasing value of $g_sN$).
The D3-brane does not need to be of Planck size and can be large when it is near the UV boundary or at time $\tau_i$. The four-dimensional spacetime here is smooth throughout with the singularity at the UV boundary arising from warping of the transverse internal space to zero-size.
Plots for the time evolution of $\rho$ and $a$ for a case of elastic collision  are shown in figure \ref{figs:cycgsp1dm3} for $g_sN=10^{-3}$ and $\delta=1$ with the time $\tau_i$ set to zero and the parameters chosen only for illustration; the D3-brane expands as it flows toward the IR and contracts as it bounces back to the UV. 

\newpage

\begin{thebibliography}{10}

\bibitem{Hubble:1929ig}
E.~Hubble, ``{A relation between distance and radial velocity among
  extra-galactic nebulae},''
{\em Proc.Nat.Acad.Sci.} {\bfseries 15} (1929) 168--173.

\bibitem{Guth:1980zm}
A.~H. Guth, ``{The Inflationary Universe: A Possible Solution to the Horizon
  and Flatness Problems},''
\href{http://dx.doi.org/10.1103/PhysRevD.23.347}{{\em Phys.Rev.} {\bfseries
  D23} (1981) 347--356}.

\bibitem{Linde:1981mu}
A.~D. Linde, ``{A New Inflationary Universe Scenario: A Possible Solution of
  the Horizon, Flatness, Homogeneity, Isotropy and Primordial Monopole
  Problems},''
\href{http://dx.doi.org/10.1016/0370-2693(82)91219-9}{{\em Phys.Lett.}
  {\bfseries B108} (1982) 389--393}.

\bibitem{Albrecht:1982wi}
A.~Albrecht and P.~J. Steinhardt, ``{Cosmology for Grand Unified Theories with
  Radiatively Induced Symmetry Breaking},''
\href{http://dx.doi.org/10.1103/PhysRevLett.48.1220}{{\em Phys.Rev.Lett.}
  {\bfseries 48} (1982) 1220--1223}.

\bibitem{Perlmutter:1998np}
{\bfseries Supernova Cosmology Project} Collaboration, S.~Perlmutter {\em
  et~al.}, ``{Measurements of Omega and Lambda from 42 high redshift
  supernovae},'' \href{http://dx.doi.org/10.1086/307221}{{\em Astrophys.J.}
  {\bfseries 517} (1999) 565--586},
\href{http://arxiv.org/abs/astro-ph/9812133}{{\ttfamily arXiv:astro-ph/9812133
  [astro-ph]}}.

\bibitem{Dvali:1998pa}
G.~Dvali and S.~H. Tye, ``{Brane inflation},''
  \href{http://dx.doi.org/10.1016/S0370-2693(99)00132-X}{{\em Phys.Lett.}
  {\bfseries B450} (1999) 72--82},
\href{http://arxiv.org/abs/hep-ph/9812483}{{\ttfamily arXiv:hep-ph/9812483
  [hep-ph]}}.

\bibitem{Kachru:2003aw}
S.~Kachru, R.~Kallosh, A.~D. Linde, and S.~P. Trivedi, ``{De Sitter vacua in
  string theory},'' \href{http://dx.doi.org/10.1103/PhysRevD.68.046005}{{\em
  Phys.Rev.} {\bfseries D68} (2003) 046005},
\href{http://arxiv.org/abs/hep-th/0301240}{{\ttfamily arXiv:hep-th/0301240
  [hep-th]}}.

\bibitem{Hailu:2011pn}
G.~Hailu, ``{Gravity Dual to Pure N=1 SU(N) Gauge Theory},''
  \href{http://dx.doi.org/10.1103/PhysRevD.84.106008}{{\em Phys. Rev.}
  {\bfseries D84} (2011) 106008},
\href{http://arxiv.org/abs/1107.6033}{{\ttfamily arXiv:1107.6033 [hep-th]}}.

\bibitem{Hailu:2011kp}
G.~Hailu, ``{Linear Confinement of Quarks from Supergravity},''
  \href{http://dx.doi.org/10.1103/PhysRevD.84.106008}{{\em Phys. Rev.}
  {\bfseries D84} (2011) 106008},
\href{http://arxiv.org/abs/1107.5827}{{\ttfamily arXiv:1107.5827 [hep-th]}}.

\bibitem{Hailu:2012jc}
G.~Hailu, ``{Glueball Masses from Linearly Confining Supergravity},''
  \href{http://dx.doi.org/10.1103/PhysRevD.85.106008}{{\em Phys.Rev.}
  {\bfseries D85} (2012) 106008},
\href{http://arxiv.org/abs/1202.2346}{{\ttfamily arXiv:1202.2346 [hep-ph]}}.

\bibitem{Kachru:2003sx}
S.~Kachru, R.~Kallosh, A.~D. Linde, J.~M. Maldacena, L.~P. McAllister, {\em
  et~al.}, ``{Towards inflation in string theory},''
  \href{http://dx.doi.org/10.1088/1475-7516/2003/10/013}{{\em JCAP} {\bfseries
  0310} (2003) 013},
\href{http://arxiv.org/abs/hep-th/0308055}{{\ttfamily arXiv:hep-th/0308055
  [hep-th]}}.

\bibitem{Silverstein:2003hf}
E.~Silverstein and D.~Tong, ``{Scalar speed limits and cosmology: Acceleration
  from D-cceleration},''
  \href{http://dx.doi.org/10.1103/PhysRevD.70.103505}{{\em Phys.Rev.}
  {\bfseries D70} (2004) 103505},
\href{http://arxiv.org/abs/hep-th/0310221}{{\ttfamily arXiv:hep-th/0310221
  [hep-th]}}.

\bibitem{Alishahiha:2004eh}
M.~Alishahiha, E.~Silverstein, and D.~Tong, ``{DBI in the sky},''
  \href{http://dx.doi.org/10.1103/PhysRevD.70.123505}{{\em Phys.Rev.}
  {\bfseries D70} (2004) 123505},
\href{http://arxiv.org/abs/hep-th/0404084}{{\ttfamily arXiv:hep-th/0404084
  [hep-th]}}.

\bibitem{Chen:2005fe}
X.~Chen, ``{Running non-Gaussianities in DBI inflation},''
  \href{http://dx.doi.org/10.1103/PhysRevD.72.123518}{{\em Phys.Rev.}
  {\bfseries D72} (2005) 123518},
\href{http://arxiv.org/abs/astro-ph/0507053}{{\ttfamily arXiv:astro-ph/0507053
  [astro-ph]}}.

\bibitem{Shandera:2006ax}
S.~E. Shandera and S.-H.~H. Tye, ``{Observing brane inflation},''
  \href{http://dx.doi.org/10.1088/1475-7516/2006/05/007}{{\em JCAP} {\bfseries
  0605} (2006) 007},
\href{http://arxiv.org/abs/hep-th/0601099}{{\ttfamily arXiv:hep-th/0601099
  [hep-th]}}.

\bibitem{Kecskemeti:2006cg}
S.~Kecskemeti, J.~Maiden, G.~Shiu, and B.~Underwood, ``{DBI Inflation in the
  Tip Region of a Warped Throat},''
  \href{http://dx.doi.org/10.1088/1126-6708/2006/09/076}{{\em JHEP} {\bfseries
  0609} (2006) 076},
\href{http://arxiv.org/abs/hep-th/0605189}{{\ttfamily arXiv:hep-th/0605189
  [hep-th]}}.

\bibitem{Hailu:2006uj}
G.~Hailu and S.~H.~H. Tye, ``Structures in the gauge / gravity duality
  cascade,'' {\em JHEP} {\bfseries 08} (2007) 009,
\href{http://arxiv.org/abs/hep-th/0611353}{{\ttfamily hep-th/0611353}}.

\bibitem{Bean:2007eh}
R.~Bean, X.~Chen, H.~Peiris, and J.~Xu, ``{Comparing Infrared Dirac-Born-Infeld
  Brane Inflation to Observations},''
  \href{http://dx.doi.org/10.1103/PhysRevD.77.023527}{{\em Phys.Rev.}
  {\bfseries D77} (2008) 023527},
\href{http://arxiv.org/abs/0710.1812}{{\ttfamily arXiv:0710.1812 [hep-th]}}.

\bibitem{Bean:2008na}
R.~Bean, X.~Chen, G.~Hailu, S.-H.~H. Tye, and J.~Xu, ``{Duality Cascade in
  Brane Inflation},''
  \href{http://dx.doi.org/10.1088/1475-7516/2008/03/026}{{\em JCAP} {\bfseries
  0803} (2008) 026},
\href{http://arxiv.org/abs/0802.0491}{{\ttfamily arXiv:0802.0491 [hep-th]}}.

\bibitem{Firouzjahi:2010ga}
H.~Firouzjahi and S.~Khoeini-Moghaddam, ``{Fields Annihilation and Particles
  Creation in DBI inflation},''
  \href{http://dx.doi.org/10.1088/1475-7516/2011/02/012}{{\em JCAP} {\bfseries
  1102} (2011) 012},
\href{http://arxiv.org/abs/1011.4500}{{\ttfamily arXiv:1011.4500 [hep-th]}}.

\bibitem{Miranda:2012rm}
V.~Miranda, W.~Hu, and P.~Adshead, ``{Warp Features in DBI Inflation},''
\href{http://arxiv.org/abs/1207.2186}{{\ttfamily arXiv:1207.2186
  [astro-ph.CO]}}.

\bibitem{Jarosik:2010iu}
N.~Jarosik, C.~Bennett, J.~Dunkley, B.~Gold, M.~Greason, {\em et~al.},
  ``{Seven-Year Wilkinson Microwave Anisotropy Probe (WMAP) Observations: Sky
  Maps, Systematic Errors, and Basic Results},''
  \href{http://dx.doi.org/10.1088/0067-0049/192/2/14}{{\em Astrophys.J.Suppl.}
  {\bfseries 192} (2011) 14},
\href{http://arxiv.org/abs/1001.4744}{{\ttfamily arXiv:1001.4744
  [astro-ph.CO]}}.

\bibitem{Liddle:2000cg}
A.~R. Liddle and D.~Lyth, {\em {Cosmological inflation and large scale
  structure}}.
\newblock Cambridge University Press,
2000.
\newblock

\bibitem{Lucini:2010nv}
B.~Lucini, A.~Rago, and E.~Rinaldi, ``{Glueball masses in the large N limit},''
  \href{http://dx.doi.org/10.1007/JHEP08(2010)119}{{\em JHEP} {\bfseries 08}
  (2010) 119},
\href{http://arxiv.org/abs/1007.3879}{{\ttfamily arXiv:1007.3879 [hep-lat]}}.

\bibitem{Khoury:2001wf}
J.~Khoury, B.~A. Ovrut, P.~J. Steinhardt, and N.~Turok, ``{The Ekpyrotic
  universe: Colliding branes and the origin of the hot big bang},''
  \href{http://dx.doi.org/10.1103/PhysRevD.64.123522}{{\em Phys.Rev.}
  {\bfseries D64} (2001) 123522},
\href{http://arxiv.org/abs/hep-th/0103239}{{\ttfamily arXiv:hep-th/0103239
  [hep-th]}}.

\bibitem{Steinhardt:2001st}
P.~J. Steinhardt and N.~Turok, ``{Cosmic evolution in a cyclic universe},''
  \href{http://dx.doi.org/10.1103/PhysRevD.65.126003}{{\em Phys.Rev.}
  {\bfseries D65} (2002) 126003},
\href{http://arxiv.org/abs/hep-th/0111098}{{\ttfamily arXiv:hep-th/0111098
  [hep-th]}}.

\end{thebibliography}

\providecommand{\href}[2]{#2}\begingroup\raggedright\endgroup

\end{document}